\documentclass[conference]{IEEEtran}
\IEEEoverridecommandlockouts
\usepackage{cite}
\usepackage{amsmath,amssymb,amsfonts}
\usepackage{algorithmic,algorithm}
\usepackage{graphicx}
\usepackage{textcomp}
\usepackage{xcolor}
\usepackage{pgfplots}
\usepackage{pgfplotstable}
\usepackage{bm}
\usepackage{supertabular}
\usepackage{caption}
\usepackage{subcaption}
\usepackage{url}
\usepackage{tikz}
\usetikzlibrary{automata, positioning, arrows}
\usepackage{diagbox}

\def  \codebook{\Phi}

\def\thard{ t^{b,*}_i }

\def\mod{\text{mod}}

\def\C{\mathbb{C}}

\begin{document}

\def\P{ \mathbb{P} }

\def\Hmin{ H_{\min} }
\def\Cmin{ C_{\min} }
\newcommand{\krd}[1]{\textcolor{violet}{ [KRD: #1] \normalsize }}
\newcommand{\mg}[1]{\textcolor{blue}{ [MG: #1 ] \normalsize }}

\def\LR{\text{LR}}
\def\LLR{\text{LLR}}
\def\rLLR{L}  
\newcommand{\absLLR}[1]{|\LLR(Y_{#1})|}
\newcommand{\absorbLLR}[1]{L_{#1}}

\makeatletter
\newcommand{\linebreakand}{%
  \end{@IEEEauthorhalign}
  \hfill\mbox{}\par
  \mbox{}\hfill\begin{@IEEEauthorhalign}
}
\makeatother

\title{
Decoding in the presence of ISI without interleaving -- ORBGRAND-AI
}

\author{
\IEEEauthorblockN{Ken R. Duffy}
\IEEEauthorblockA{\textit{Dept. of ECE \& Dept. Mathematics} \\
\textit{Northeastern University}\\
Boston, USA \\
k.duffy@northeastern.edu}
\and
\IEEEauthorblockN{Moritz Grundei}
\IEEEauthorblockA{\textit{Electrical and Computer Engineering} \\
\textit{Technical University of Munich}\\
Munich, Germany \\
moritz.grundei@tum.de}
\and
\IEEEauthorblockN{Jane A. Millward}
\IEEEauthorblockA{\textit{Research Laboratory of Electronics} \\
\textit{Massachusetts Institute of Technology}\\
Cambridge, USA \\
janem7@mit.edu}
\linebreakand
\IEEEauthorblockN{Muralidhar Rangaswamy}
\IEEEauthorblockA{\textit{Sensors Directorate} \\
\textit{Air Force Research Laboratory}\\
WPAFB, USA \\
muralidhar.rangaswamy@us.af.mil}
\and
\IEEEauthorblockN{Muriel M\'edard}
\IEEEauthorblockA{\textit{Research Laboratory of Electronics} \\
\textit{Massachusetts Institute of Technology}\\
Cambridge, USA \\
medard@mit.edu}
}

\maketitle

\begin{abstract}
Inter symbol interference (ISI), which occurs in a wide variety of channels, is a result of time dispersion. It can be mitigated by equalization, which results in noise coloring. Inspired by the development of Approximate Independence in statistical physics, for such colored noise we propose a decoder called Ordered Reliability Bits Guessing Random Additive Noise Decoding (ORBGRAND-AI) that operates without the need for turbo equalization or interleaving. By foregoing interleaving, ORBGRAND-AI can deliver the same, or lower, block error rate (BLER) for the same amount of energy per information bit in an ISI channel as a state-of-the-art soft input decoder, such as Cyclic Redundancy Check Assisted-Successive Cancellation List (CA-SCL) decoding, with an interleaver. To assess the decoding performance of ORBGRAND-AI, we consider delay tap models and their associated colored noise. In particular, we examine a two-tap dicode ISI channel as well as an ISI channel derived from data from RFView, a physics-informed modeling and simulation tool. We investigate the dicode and RFView channel under a variety of imperfect channel state information assumptions and show that a second order autoregressive model adequately represents the RFView channel effect. 
\end{abstract}


\section{Introduction}
\let\thefootnote\relax\footnotetext{
Preliminary versions of this paper were presented in the 2023 Globecom, 2024 SPAWC, and Asilomar conferences \cite{duffy23using,millward2024using,millward2024enhancing}. Due to space limitations, those papers made succinct observations about the impact of correlated ISI on decoder performance. This paper extends the work of the conference papers by providing a more well-rounded treatment of the problem, via an explicit demonstration of the fact that the entropy of the correlated ISI channel is less than that of the uncorrelated ISI channel. Furthermore, the rationale for approximating the interference generated using RFVIEW as an AR(2) process is established. The simulations are substantially more comprehensive compared to the conference submissions.}

In many modern communication systems, inter-symbol interference (ISI) is a key challenge that must be addressed to ensure efficient and reliable transmission \cite{proakis2008digital}. ISI is for example caused by multipath propagation in wireless environments, where multiple delayed and attenuated copies of the transmitted signal arrive at the receiver \cite{rappaport2002wireless}. A common approach to mitigate ISI is the application of equalization techniques such as zero-forcing (ZF) or minimum mean-squared error (MMSE) filtering. Although these techniques effectively suppress ISI, they necessarily introduce noise correlation, resulting in noise components that are statistically dependent across symbols.

Most soft input forward error correction decoders, however, 
assume noise is independent across symbols. It has been observed that the performance of such decoders degrades significantly in the presence of correlated noise \cite{lin2004error}. One widely deployed remedy for correlated noise in signal-processing chains is the use of interleavers. Interleavers collect a batch of coded blocks and transmit a shuffled ordering of the encoded symbols to the receiver. Once the full batch has been received, the receiver applies the inverse permutation (de-interleaving), thereby reconstructing the original coded blocks, which are then passed to the decoder in their original order. 

If the batch size is chosen to be sufficiently large, this temporal permutation can decorrelate the effective noise so that, from the decoder's perspective, the disturbances within each reconstructed block are closer to being independent. This benefit comes at the cost of increased end-to-end latency, since the receiver must buffer an entire interleaving batch before de-interleaving and decoding can proceed \cite{sklar2001digital}. Moreover, from an information-theoretic perspective, correlated noise has lower entropy than uncorrelated noise with the same marginal statistics as statistical dependence reduces overall uncertainty, and so channel capacity is higher without interleaving \cite{Gallager68,Cover91}.

While interleaving is widely used and provides an effective means of matching transformed channel conditions to the modeling assumptions underlying most error-correcting decoders, it has been shown that noise correlation can be exploited to improve decoding performance for some decoders. 
Most prominently, Douillard et al. introduced turbo equalization, an iterative scheme that exchanges soft information between an equalizer and a soft-input soft-output (SISO) decoder to explicitly leverage channel-induced correlation \cite{douillard1995iterative, koetter2004turbo}. Although incurring a complexity and latency cost due to the iterative nature of the algorithm, the approach is effective for codes that possess an efficient SISO decoder.
Here we explore a mechanism through which arbitrary moderate-redundancy block codes can be decoded in the presence of correlated noise without the need for turbo equalization.

GRAND (Guessing Random Additive Noise Decoding) is a noise-centric decoding paradigm that can be applied to any moderate or low redundancy block code. It operates by querying candidate noise sequences in approximately decreasing order of likelihood. In additive-noise settings, this procedure realizes maximum or near maximum-likelihood (ML) decoding performance in both soft and hard detection settings \cite{duffy19GRAND,solomon20SGRAND,duffy22ORBGRAND,liu2022orbgrand}.  Owing to its noise-centric approach, GRAND can directly exploit structure in discrete channels, for example by incorporating Markovian correlation models into the noise-guessing process to improve decoding performance \cite{An22,rezaei2023coding}. In soft-detection settings, however, capturing the potential gains from noise correlation in a practical way requires distinct innovation. 

By adopting techniques from thermodynamic probability theory to manage dependence and combining them with a symbol level noise querying order, we demonstrate that it is possible to accurately soft detection decode any moderate redundancy code with an arbitrary noise correlation pattern without the need for turbo equalization or interleaving. The approach relies on the principle of Approximate Independence (AI), which treats suitably defined blocks of symbols as approximately independent while explicitly accounting for correlations within small symbol neighborhoods. By incorporating correlation information directly into the noise-guessing order, decoding performance can be improved without the need for turbo equalization, while simultaneously reducing the complexity and latency typically introduced by interleaving. This makes the approach particularly attractive for ultra-reliable low-latency communication (URLLC) scenarios.

The contributions of this paper are twofold. First, we introduce ORBGRAND-AI, a decoder that, independently of the underlying code structure, can exploit arbitrary correlation in additive noise caused by equalizing ISI to inform and improve decoding. In contrast to common assumptions that correlation must be removed prior to decoding, ORBGRAND-AI leverages it directly within the noise-querying process, yielding improvements of several dB in both block error rate (BLER) and bit error rate (BER). Additionally, we provide heuristic arguments to explain the origin of these gains.

Second, we demonstrate that these gains persist under realistic channel conditions. In particular, we consider channel models in which noise coloring arises from practical filtering operations, and scenarios involving higher-order modulation schemes. We further evaluate performance under adverse conditions such as quantization effects and channel estimation inaccuracies, showing that the proposed approach remains robust in these settings.

This paper is structured as follows. Section~\ref{sec:chan} introduces the two channel models considered in this work: the dicode channel and the RFView channel. In Section~\ref{sec:capacity}, we provide a heuristic argument explaining why treating small blocks of symbols as approximately independent can yield significant gains in correlated channels. Section~\ref{sec:orbgrand_ai} presents a detailed description of ORBGRAND-AI. 
Section~\ref{sec:perfeval} evaluates the performance of ORBGRAND-AI and benchmarks it against reference decoders, including CA-SCL and the original ORBGRAND. Finally, Section~\ref{sec:discussion} analyzes the performance of ORBGRAND-AI in both the dicode and RFView channels across a range of operating conditions.

\begin{figure*}
    \centering
    \begin{tikzpicture}[node distance=0.5cm]

        \node[state](S_P) [rectangle]{S/P};
        \coordinate[left=of S_P] (orig);
        \node[state](enc_2) [rectangle, right=of S_P]   {encoder};
        \node[state](enc_1) [rectangle, above=of enc_2] {encoder};
        \node[state](enc_3) [rectangle, below=of enc_2] {encoder};
        \node[state](inter) [rectangle, right=of enc_2] {interleaver};
        \node[state](mod)   [rectangle, right=of inter] {modulator};
        \node[state](chan)  [rectangle, right=of mod]   {channel};
        \node[state](equal) [rectangle, right=of chan]  {equalizer};
        \node[state](demod) [rectangle, right=of equal] {demodulator};
        \node[state](deint) [rectangle, right=of demod] {deinterleaver};
        \node[state](dec_2) [rectangle, right=of deint] {decoder};
        \coordinate[right=of dec_2](dest_2);
        \node[state](dec_1) [rectangle, above=of dec_2] {decoder};
        \coordinate[right=of dec_1](dest_1);
        \node[state](dec_3) [rectangle, below=of dec_2] {decoder};
        \coordinate[right=of dec_3](dest_3);
        \node[text width=0.5cm, rotate=90] [below=of enc_1, xshift=0.3cm, yshift=0.1cm] {\normalsize $\hdots$};
        \node[text width=0.5cm, rotate=90] [below=of enc_2, xshift=0.3cm, yshift=0.1cm] {\normalsize $\hdots$};
        \node[text width=0.5cm, rotate=90] [below=of dec_1, xshift=0.3cm, yshift=0.1cm] {\normalsize $\hdots$};
        \node[text width=0.5cm, rotate=90] [below=of dec_2, xshift=0.3cm, yshift=0.1cm] {\normalsize $\hdots$};
    
        \path[->] 
                (orig)  edge              node                     {} (S_P)
                (S_P)   edge              node                     {} (enc_1)
                        edge              node                     {} (enc_2)
                        edge              node                     {} (enc_3)
                (enc_2) edge              node                     {} (inter)
                (enc_1) edge              node                     {} (inter)
                (enc_3) edge              node                     {} (inter)
                (inter) edge              node                     {} (mod)
                (mod)   edge              node                     {} (chan)
                (chan)  edge              node                     {} (equal)
                (equal) edge              node                     {} (demod)
                (demod) edge              node                     {} (deint)
                (deint) edge              node                     {} (dec_1)
                        edge              node                     {} (dec_2)
                        edge              node                     {} (dec_3)
                (dec_1) edge              node                     {} (dest_1)
                (dec_2) edge              node                     {} (dest_2)
                (dec_3) edge              node                     {} (dest_3);

    \end{tikzpicture}
    \caption{Signal processing chain of a bit interleaved communication system}
    \label{fig:system}
\end{figure*}
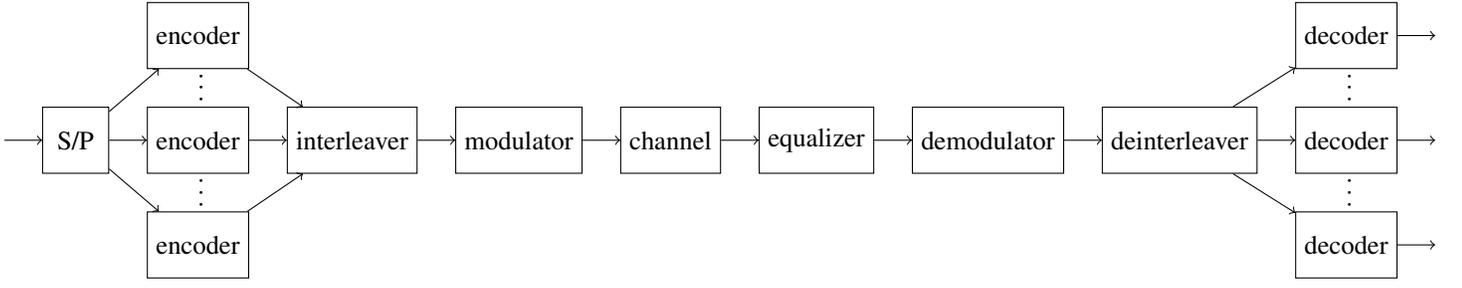

\section{ISI Channel Description}
\label{sec:chan}
The standard structure of an interleaved communication system is shown in Fig. \ref{fig:system}. 
For the channel, we consider a linear model 
\begin{align*}
Y_{k'} = \sum_{j\geq 0 }h_{k',j}X_{k'-j} + N_{k'},   
\end{align*}
where $k'\in\mathbb{Z}^+$ denotes the symbol time scale, $X_{k'}$ is the complex-valued transmitted symbol, $N_{k'}$ is complex white Gaussian noise, and $h_{k'}$ denotes the discrete representation of the channel impulse response at the time of the transmission. We use $k'$ as we use $k$ later to denote the information bits in a codeword as per convention. ISI is generally the result of time dispersion imparted on the transmitted signal. We shall denote, unless otherwise stated, random variables by capital letters and sample values or constants by lower case letters. Vectors and matrices we shall denote by underlining. We denote the dimensions of vectors and matrices using superscripts and specific coordinates within the matrices and vectors with subscripts.

When we consider a sequence of symbols of length $n_s$, our model becomes: 
\begin{equation} \label{eq:rfview_model}
   \underline{Y}^{n_s} = \underline{h}^{n_s\times n_s} \underline{X}^{n_s} + \underline{N}^{n_s}
\end{equation}

\noindent where $\underline{Y}^{n_s} = (Y_1, Y_2, ..., Y_{n_s})^T$ denotes the sequence of transmitted symbols, $\underline{N}^{n_s}$ denotes a vector of additive white Gaussian noise with auto-covariance $\underline{C}^{n_s\times n_s}_{N} = \sigma_N^2 \underline{I}^{n_s\times n_s}$ and $\underline{h}^{n_s\times n_s}$ is the channel matrix whose elements are the $h_{k',j}$s. We denote the identity matrix of size $n_s\times n_s$ by $\underline{I}^{n_s\times n_s}$. When we consider codewords of length $n$, we denote the vectors corresponding to the received code word and noise effect by $\underline{Y}^n$, $\underline{X}^n$ and $\underline{N}^n$ respectively.

\subsection{Role of Interleavers}

Most modern communication systems rely on an inner soft-decision forward error correction code with a short interleaver to handle fast fading \cite{etsi138212,elfouly2022burst}, and an outer code with a deep interleaver to handle longer outages \cite{ccsds1310b5}. Fine-grained interleaving helps eliminate localised burst errors that would be damaging to forward error correction decoders \cite{elfouly2022burst}. Deeper interleaving assists the outer code overcome the loss of large contiguous data blocks due to erasures or inner code decoding failures \cite{ccsds1310b5}.

For bit-interleaved coded modulation (BICM) and related architectures, interleaving is used to bring the effective channel seen by the demapper and decoder closer to a memoryless model, thereby reducing the mismatch between the true channel statistics and the assumptions built into conventional soft demappers and decoders \cite{caire1998bicm,martinez2009bicm,Li24ORBGRANDBICM}. Here, interleaving spreads temporally localized error events over multiple codewords so that they can be handled more effectively by the underlying error-correcting code \cite{proakis2008digital,shi2004interleaving,elfouly2022burst}. This role is particularly relevant in equalized ISI channels, where ISI suppression often introduces residual correlation in the post-equalization noise. From the perspective of a conventional inner decoder, interleaving disperses this dependence across codewords and makes the effective observations appear closer to independent. The second role of interleaving is channel outage mitigation. Outage errors occur when entire blocks of data are not received or when the inner decoder announces failed decoding. Deeper interleaving enables the application of outer codes to overcome this challenge, e.g. \cite{ccsds1310b5,lun2008packetnetworks,medard2025network}. The scope of the present work is focused on the first of these two roles, which, in turn, assists the second.

The purpose of ORBGRAND-AI is to show that the use of interleaving to enforce decoder-friendly independence assumptions after equalization can be reduced or avoided when the decoder itself explicitly accounts for channel-induced correlation. In other words, the contribution of ORBGRAND-AI is to mitigate channel--decoder mismatch at the physical layer by exploiting the residual noise structure directly during decoding without the need for turbo equalization.

\subsection{Delay Tap Channel Model} 

We use a delay line model to generate the channel profile in terms of its paths indexed by $d$ ranging from $1$ to $p_{k', j}$ where $k'$ denotes the symbol time index and $j$ denotes the delay due to ISI. Each path $d$ has complex attenuation $\{a_{k',j, d}\}$ and delays $\{\tau_{k',j,d}\}$. Given a sampling frequency $f_s$ we can compute the time discrete channel impulse response: $h_{k',j} = \sum_{d= 1}^{p_{k', j}} a_{k',j, d}\text{sinc}\left(\tau_{k',j, d} f_s-k'\right)$. A representative value for delay spread, defined as the maximum difference among the $\tau_d$s is $1\mu s$ for terrestrial outdoor systems.

A special case of the delay tap channel model is the dicode partial response channel. It is a simple example of ISI as there are only two channel taps: 
\begin{equation*}
    h_{k',j} = \begin{cases}
                1 \hspace{0.3cm}&j=0,\\
                -\rho           &j=1,\\
                0               &\text{otherwise},
          \end{cases} 
\end{equation*}
with $\rho\in[0,1]$. 

Equalization through zero forcing removes ISI but leads to colored noise, $\{\Tilde{N}_{k'}\}$, with an autoregressive description $\Tilde{N}_{k'} = \rho\Tilde{N}_{k'} +  N_{k'}$ where $N_{k'}$ denotes the Gaussian noise prior to equalization. This type of colored noise is commonly referred to as Gauss-Markov noise and it exhibits exponentially decaying temporal correlation strength: $\mathbb{E}[|\Tilde{N}_{k'} \Tilde{N}_{i'}|] \propto\rho^{|k'-i'|}$ where $i' \in \mathbb{Z}^+$ is another variable denoting the symbol time scale (we use $i$ later to denote the block index when we discuss ORBGRAND-AI). 

\subsection{RFView ISI Channel} \label{sec:rfview_dataset_channel}

We consider an ISI channel generated using channel impulse response data from RFView. RFView is a high-fidelity, physics-based RF simulation and modelling tool \cite{gogineni2022high}. The RFView dataset consists of the in-phase and quadrature (I-Q) channel clutter impulse response of a mixed terrain environment with some discrete clutter sources, like buildings \cite{gogineni2022high}. The dataset contains 30 coherent processing intervals (CPIs). Each CPI consists of a 3D data cube comprised of 32 antenna channels, 64 pulses and 2335 impulse response samples sampled at $10$ MHz (refer to Fig.~\ref{fig:rfview_visualization}). 
\begin{figure}[htbp]
    \centering
      \begin{subfigure}[b]{0.48\textwidth}
         \centering
         \includegraphics[width=\textwidth, trim={7cm 8cm 8cm 6.5cm},clip]{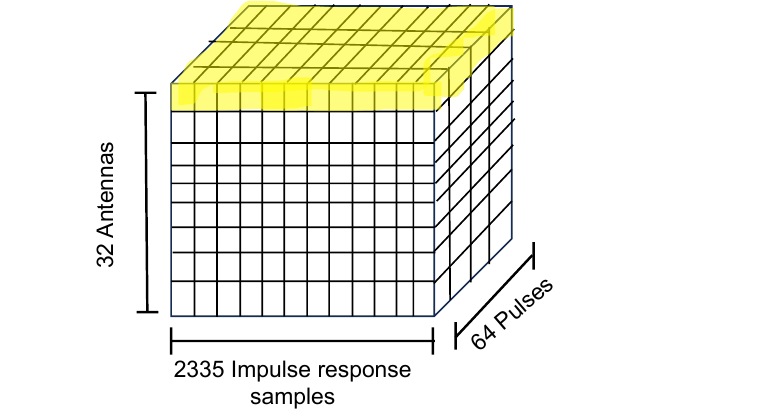}
      \end{subfigure}
    \caption{Illustration of RFView dataset for a single CPI. We process the data from each data cube corresponding to a particular CPI for the first antenna element only as we consider single-input single-output communications scenarios. }
    \label{fig:rfview_visualization}
\end{figure}

To provide a channel estimate consistent with our earlier ISI channel definition, we process each data cube corresponding to a particular CPI. Since radar data is usually comprised of several pulses, we treat the pulse axis within each data cube as ``slow time" and we assume the channel impulse response changes from pulse to pulse (i.e the channel impulse response from time sample 1 to 2335 corresponds to the impulse response at pulse index 1 and for time samples 2336 to $2335 \times 2 = 4760$ the channel impulse response corresponds to the channel impulse response at pulse 2, etc). Additionally, we only process data from a single antenna element (the first element along the antenna dimension in Fig. ~\ref{fig:rfview_visualization}) as we only consider single-input single-output in our initial investigations. 

To obtain the estimate of each coefficient $h_{k',j}$, we transmit a complex passband pulse through the channel impulse response corresponding to a particular pulse index. We set $f_s = 10$MHz and the length of the pulse to $L = 467$ which ensures that we obtain a 6-tap ISI channel i.e $j \in \{1,..., 6\} \forall k'$ for the RFView channel. We set the carrier frequency, $f_c$, of the complex passband sounding signal to $\frac{f_S}{L}$. We denote the complex passband pulse by $u_l$ where
\begin{equation*}
    u_l = \begin{cases}
        a \exp(i 2\pi f_c l) & l = 1,\ldots, L, \\
        0 & \text{ otherwise}.
    \end{cases}
\end{equation*}
The constant $a$ is selected so that$\sum_{l=1}^{L} |u_l|^2 = 1$ and we denote the full vector of the sounding signal by $\underline{u}^L$.   

The channel response given by RFView is $\underline{g}^{m, \mu}$, which is a matrix that takes into account the $m = 2335$ impulse response samples  sampled at $10$MHz  and the $\mu = 64$ pulses. For each $r \in \{1,...\mu\}$, we  are able to transmit five sounding signals per $\underline{g}^{m, r}$ and a total of $5 \times 64 = 320$ sounding signals per CPI. For each $\underline{g}^{m, r}$, we can transmit five sounding signals since $\frac{m}{L} = 5$. We transmit each sounding signal through the channel separately so that we are able to measure the individual effect of each sounding signal and therefore construct the 6-tap ISI channel. We set $r(k') = \frac{k'+(5-k' \mod{5} )}{5}$ to account for the fact that we are able to transmit 5 sounding signals per $\underline{g}^{m,r(k')}$ but wish to isolate the output response of each sounding signal individually to be able to construct the ISI coefficients. We account for the fact that each sounding signal will be delayed by L samples more than the previous one later when we construct the coefficients, $h_{k',j}$ by sampling from the matched filter output of the channel. From a data processing perspective we do not need to account for the delays in the sounding signals because adding the delay, effectively zero-padding, does not affect the result of the channel output response convolutions. We account for the delay later when we go to construct $\underline{h}^{n_s\times n_s}_{RFV}$.

The noiseless output $\underline{z}^{\zeta, 320}$ of the $(k') ^{th}$ sounding signal $k'\in \{1,...,320\}$ and the channel response at pulse index $r(k')=\frac{k'+(5-k' \mod{5} )}{5}$ is given by the convolution $\underline{z}_{k'} = \underline{g}^{m, r(k')} * \underline{u}^L$ with components given
\begin{equation*}
    z_{q, k'} = \sum_{\kappa=1}^{m}  g_{\kappa, r(k')} u_{q - \kappa}
\end{equation*}
We have by the properties of convolution that $\zeta = m + L - 1$. 

We next perform matched filtering on $\underline{u}^L$ to obtain $
\underline{z}'_{k'} = \underline{z}_{k'} * (\underline{u}^{L})'$ for each sounding signal $k'$ where $(\underline{u}^{L})'$  denotes the matched filter response of $\underline{u}^L$ as defined in \cite{anderson2005optimal}. $\underline{z}'_{k'}$ has components
\begin{equation}
    z'_{q, k'} = \sum_{\kappa = 1}^\zeta z_{\kappa, k'} u'_{q - \kappa}
\end{equation}
\noindent The full matrix of the matched filter output is $\underline{z}'^{\eta, 5\mu}$ where $\eta$ is $ 2L + m - 2 = 2*467 + 2335 - 2 = 3267$. 

We next sample the elements of $\underline{z}'^{\eta, 5\mu}$ at intervals of L along the impulse response axis (i.e the axis with dimension $\eta$) to obtain $\underline{z}''^{6, 5\mu}$. The $\eta$ axis has now been reduced to a dimension of 6 since $\lfloor{\frac{\eta}{L}}\rfloor = 6$. The samples of $\underline{z}''^{6, 5\mu}$ become the ISI coefficients $h_{k',j}$ for $j \in \{1,...,6\}$ via
\begin{equation*}
    h_{k',j} = \underline{z}''_{j, k'-(j-1)}.
\end{equation*}
We next construct the matrix $\underline{h}^{n_s\times n_s}_{RFV}$ using the $h_{k',j}$s obtained above by: 
\begin{equation*}
    \underline{h}^{n_s\times n_s}_{RFV} = \begin{bmatrix}
        h_{1, 1} & 0 & 0 & 0 &\dots & 0\\
        h_{2, 2} & h_{2, 1} & 0 & 0 & \dots & 0\\
        h_{3, 3} & h_{3, 2} & h_{3, 1} & 0& \dots &0\\
        \vdots & \vdots & \ddots & \vdots & \vdots & \vdots\\
        0 & \dots & 0 &  h_{n_s, 6} & \dots & h_{n_s, 1}
    \end{bmatrix}.
\end{equation*}
We complete this process for each of the 30 CPIs. In our simulations, we uniformly sample from these 30 matrices to obtain the channel realization $\underline{h}^{n_s\times n_s}_{RFV}$.

\section{A Theoretical Heuristic}
\label{sec:capacity}
We will now argue through the example of an equalized dicode channel (or equivalently Gauss-Markov noise) that there is a significant gain to be realized when we only consider correlations across small neighborhoods (blocks) of received symbols and treat the blocks themselves as independent with regard to one another.

With variance $\sigma_N^2$ and correlation coefficient $\rho\in(0,1]$, assume that the continuous noise sequence $\{N^{n_s}\}$ is a zero-mean complex-valued Gaussian with auto-covariance matrix $\underline{C}_{N}^{n_s\times n_s}\in \mathbb{R}^{n_s\times n_s}$ having entries $\underline{C}_{N_{k',i'}} = \sigma_N^2\rho^{|k'-i'|}$. The normalized differential entropy rate of $\underline{N}^{n_s}$ can be calculated as 
\begin{align*}
          & \log(2e\pi) + \dfrac{1}{n}\log(|\underline{C}_{N}^{n_s\times n_s}|)  \nonumber\\
          & = \log(2e\pi\sigma_N^2) + \left(1- \dfrac{1}{n_s}\right) \log(1-\rho^2),
          \label{eq:GMentropy}
\end{align*}
e.g. eq. (9.34) \cite{Cover91}. The final term encapsulates the decrease in entropy that arises from channel correlation as $\log(1-\rho^2)<0$ for $\rho>0$. In a heavily interleaved channel $\rho=0$ and the final term is zero. If the channel was truly independent for each block of $b$ bits, then $C_{N_{k',i'}}$ would be $0$ for $|k'-i'|>b$ and the normalized differential entropy rate would instead be
\begin{align*}
           \log(2e\pi\sigma_N^{2}) + \left(1- \dfrac{1}{b}\right) \log(1-\rho^2),
\end{align*}
where the only difference is the multiplier of the final term, which has changed from $(1-1/n_s)$ to $(1-1/b)$. Thus, in this setting, to get more than half of the reduction in normalized differential entropy, a block-size of $b=2$ suffices, suggesting significant gains should be available with small blocks.

The principle of treating neighboring blocks as approximately independent random variables originates from considerations in thermodynamic probability theory where stochastic processes are approximated by product measures across boundaries \cite{Sullivan1976,Lewis95B,Pfister02,Pfister04B}. 

We now show that we can expect similar performance gains in the case of a second-order Gauss-Markov process. We analyze the entropy rate of a second-order Gauss-Markov process because, as per Burg's theorem, the entropy of a stochastic-process subject to $\alpha \in \mathbb{Z}^+$ covariance constraints is maximized by a $\alpha$-th order Gauss-Markov process \cite{choi1984information}.

We proceed by calculating the normalized differential entropy rate of a second-order Gauss-Markov process which has the following covariance constraints: 
\begin{align*}
    \mathbb{E}[N_{k'}  N_{k'}] &= \sigma_N^2\\
    \mathbb{E}[N_{k'}  N_{k'+1}] &= \rho_1 \sigma_N^2 \\
    \mathbb{E}[N_{k'}  N_{k'+2}] &= \rho_2 \sigma_N^2 \\
\end{align*}
for  $k' = 1, 2, 3...$. We note that $\rho_1$ and $\rho_2$ denote the correlation coefficients and in our physical model $\rho_1, \rho_2 \in (0, 1]$. For $i'>2$, the cross-covariance terms are
\begin{equation*}
     \mathbb{E}[N_{k'} N_{k'+i'}] = \beta_1 \mathbb{E}[N_{k'} N_{k'+i'-1}] + \beta_2 \mathbb{E}[N_{k'} N_{k'+i'-2}] 
\end{equation*}
where $\beta_1, \beta_2$ are the coefficients of the time-series and can be found by solving the Yule-Walker equations \cite{shumway2017time}. 

Using the recursive expression for $\mathbb{E}[N_{k'} N_{k'+i'}]$ in conjunction with Wolfram Mathematica, we find that the determinant of the auto-covariance matrix for $n_s \geq 4$ is
\begin{equation*}
    |\underline{C}^{n_s\times n_s}_{N}| = -\frac{(\rho_2 -1)^{n_s-2}(1-2\rho_1^2 +\rho_2)^{n_s-2}(\sigma_N^2)^{n_s}}{(\rho_1^2-1)^{n_s-3}}.
\end{equation*}
We recall that $\rho_1$ and $\rho_2$ denote the correlation coefficients which in our physical model must be numbers between 0 and 1. This means that when $n_s$ is odd $-(\rho_2 -1)^{n_s-2}$ is positive and $(\rho_1^2 -1)^{n_s-3}$ is positive, and when $n_s$ is even $-(\rho_2 -1)^{n_s-2}$ is negative and $(\rho_1^2 -1)^{n_s-3}$ is negative. Since $\underline{C}^{n_s\times n_s}_N$ is a positive semi-definite matrix, we now need to check that $1 - 2\rho_1^2 + \rho_2$ is positive (i.e $\rho_1^2 < \frac{\rho_2 + 1}{2}$) as this will ensure that the determinant is positive. 

The Yule-Walker equations impose the following conditions on the selection of $\rho_1$ and $\rho_2$ via the variance of the innovation process in the time series:
\begin{align*}
    & 0 < \frac{\rho_1 (\rho_2-1)}{\rho_1^2 -1}\rho_1 + \frac{\rho_1^2-\rho_2}{\rho_1^2 -1}\rho_2 < 1\\
    \implies & 0 < \frac{\rho_1^2 + \rho_2^2 - 2\rho_1^2 \rho_2}{1- \rho_1^2} < 1.
\end{align*}
Manipulating the  right-hand side of the variance constraint we find that
\begin{equation*}
    \rho_1^2 < \frac{\rho_2 + 1}{2},
\end{equation*}
which ensures that the $1 - 2\rho_1^2 +\rho_2$ term, in the expression for the determinant is indeed positive.

We can now write the normalized differential entropy rate for second-order Gauss-Markov noise as 

\begin{equation*}
    \frac{1}{2}\log(2\pi e\sigma_N^2) + \frac{1}{2n_s}\log \left ( -\frac{(\rho_2 -1)^{n_s-2}(1-2\rho_1^2 +\rho_2)^{n_s-2}}{(\rho_1^2-1)^{n_s-3}} \right ) 
\end{equation*}

We now proceed to use the expression we have found for the normalized differential entropy rate to find the capacity of a channel subject to second-order Gauss-Markov noise. The channel capacity can be computed  as 
\begin{equation*}
    \mathbf{C} = \sup_{X'} \underline{I}(X'; Y')
\end{equation*}

\noindent where $ \underline{I}(X'; Y')$ is the lim-inf information rate \cite{verdu1994general} with 
\begin{equation*}
    \underline{I}(X'; Y') = \liminf_{n\rightarrow \infty} \frac{1}{n} \log \frac{P_{\underline{Y}^{n_s}|\underline{X}^{n_s}}(\underline{y}^{n_s}|\underline{x}^{n_s})}{P_{\underline{Y}^{n_s}}(\underline{y}^{n_s})}
\end{equation*}
\noindent and $X', Y'$ denote sequences of the finite dimensional distributions $X' = \{\underline{X}^{n_s} = \{X_1^{(n_s)},...X_{n_s}^{(n_s)}\}\}_{n_s=1}^\infty$ and $\underline{Y}' = \{Y^{n_s} = \{Y_1^{(n_s)},...Y_{n_s}^{(n_s)}\}\}_{n_s=1}^\infty$ respectively. There is an additional inequality relating the lim-inf information rate and the lim-sup entropy rate in \cite{verdu1994general}
\begin{equation*}
     \underline{I}(X'; Y') \leq \overline{H}(Y') - \overline{H}(Y'|X')
\end{equation*}
where $\overline{H}(\cdot)$ denotes the lim-sup entropy rate defined as  
\begin{equation*}
    \overline{H}(Y') = \limsup_{n\rightarrow \infty} \frac{1}{n} \log \frac{1}{P_{\underline{Y}^{n_s}}(\underline{y}^{n_s})}. 
\end{equation*}
$\overline{H}(Y'|X')$ is defined similarly. Applying this inequality to find the capacity of second-order Gauss-Markov noise, we find that 
\begin{align*}
    &\underline{I}(X'; Y') \leq \overline{H}(Y') - \frac{1}{2}\log(2\pi e\sigma_N^2) \\
    &- \lim_{n_s\rightarrow \infty}\frac{1}{2n_s}\log \left ( -\frac{(\rho_2 -1)^{n_s-2}(1-2\rho_1^2 +\rho_2)^{n_s-2}}{(\rho_1^2-1)^{n_s-3}} \right ).
\end{align*}
\noindent Noting that $\overline{H}(Y')$ is maximized when each $\underline{Y}^{n_s}$ is a Gaussian with auto-covariance $(\sigma_X^2 + \sigma_N^2)\underline{I}^{n_s \times n_s}$ where $\sigma_X^2$ denotes the maximum power of the data symbols, we find 
\begin{align*}
    \mathbf{C} \leq & \frac{1}{2} \log(2\pi e) + \frac{1}{2}\log (\sigma_X^2 + \sigma_N^2) 
    -  \frac{1}{2}\log(2\pi e\sigma_N^2) \\
    &- \lim_{n_s\rightarrow \infty}\frac{1}{2n_s}\log \left ( -\frac{(\rho_2 -1)^{n_s-2}(1-2\rho_1^2 +\rho_2)^{n_s-2}}{(\rho_1^2-1)^{n_s-3}} \right ),
\end{align*}
where the inequality remains because of Hadamard's inequality; the auto-covariance matrix of $\underline{Y}^{n_s}$ has off-diagonal terms so we bound it using Hadamard's inequality. As the auto-covariance matrix of the second-order Gauss-Markov process, $\underline{C}_N^{n_s \times n_s}$, has off-diagonal terms, the lim-sup entropy rate of the second order Gauss-Markov noise, $\overline{H}(Y'|X')$, must be less than the lim-sup entropy rate of uncorrelated Gaussian noise as a consequence of Hadamard's inequality. This result becomes helpful when we approximate the RFView channel by a second order autoregressive process (AR(2)) later in the paper. 

\section{ORBGRAND-AI}
\label{sec:orbgrand_ai}
\subsection{GRAND}
Guessing Random Additive Noise Decoding (GRAND) is a family of codebook agnostic decoders. The central premise of GRAND is that, in additive error channels for codewords of length $n$ with $\underline{Y}^n=\underline{X}^n\oplus \underline{N}^n$, the entropy of the noise $\underline{N}^n$ is typically much smaller than the entropy of the code word $\underline{X}^n$. GRAND finds a decoding by iteratively guessing the noise realization $\underline{N}^n$ and subsequently inverting the channel until a code word is found \cite{duffy19GRAND}. In the case of a binary linear code with parity check matrix $H\in\{0,1\}^{n-k\times n}$, GRAND computes the syndrome $H(\underline{Y}^n \oplus \underline{N}^n_{g'})$ for each noise guess $\underline{N}^n_{g'}$. If the syndrome is zero, a decoding is found, else, GRAND continues guessing. Assuming that the noise sequences are queried in decreasing likelihood order given soft input or channel statistics, GRAND is maximum likelihood (ML) achieving \cite{duffy19GRAND,solomon20SGRAND}. When an approximate query order is used, GRAND is approximately ML achieving, e.g. \cite{duffy22ORBGRAND,yuan2023guessing,abbas2023listgrand,Li24RSORBGRAND}.

\subsection{ORBGRAND}
ORBGRAND is a soft detection variant of GRAND. Given an ordered list of bit reliabilities by a soft demapper, ORBGRAND uses a linear approximation of the reliability curve to turn the problem of finding a decreasing likelihood guesswork function into generating binary sequences in increasing order of logistic weight \cite{duffy22ORBGRAND}. This sequence, in addition to the reliability order, is then used to produce the noise realizations. A multi-line approximation of the reliability curve has been investigated too \cite{duffy22ORBGRAND}. During the demapping and the generation of binary query sequences in increasing order of logistic weight, ORBGRAND assumes independent bits and thus relies on interleaving. Generating sequences in increasing logistic weight order may be done using the landslide algorithm \cite{duffy22ORBGRAND,riaz2024sub}. In general, ORBGRAND is well-suited to efficient implementation in hardware, eg. \cite{abbas2022high,condo2022fixed, Riaz23, xiao2023low,riaz2024sub}. 

\subsection{Symbol-level ORBGRAND}

Symbol-level ORBGRAND \cite{An22} introduced a modulation-aware variant that assumes symbols experience independent channel effects consistent with symbol-level interleaving. Given a hard detected symbol, its neighbors in the constellation are considered as potential substitutions. The exceedance distance between potential substitution symbols and the hard detected symbol is used as a reliability input for ORBGRAND's rank ordering, whereupon the original noise effect pattern generator is employed. In contrast to bit level ORBGRAND, symbol level ORBGRAND uses the generated patterns to pick symbols to substitute. Hence, symbols with lower exceedance distance are swapped in earlier. If a symbol substitution pattern proposes a single symbol be substituted more than once, the pattern is discarded. Empirical results demonstrate that symbol-level ORBGRAND can achieve identical performance to operating on bit level reliabilities while realizing a reduction in pattern generation complexity.
\subsection{ORBGRAND-AI}
To enable a receiver to detect or correct errors, prior to transmission each collection of $k$ information bits is coded to a $n>k$ bit code-word $c^n=(c_1,\ldots,c_n)\in\{0,1\}^n$. For spectral efficiency, most communication systems employ high-order modulation where each transmitted symbol communicates multiple bits of information \cite{Proakis}. If a modulation scheme is employed with a complex constellation of size $|\chi|=2^{m_s}$, the $n$ coded bits are translated into $n_s=n/m_s$ symbols by sequentially mapping each collection of $m_s$ bits to the corresponding higher order symbol. In the absence of interleaving, this results in the transmission of the higher order sequence $\mod(c^n) = \underline{X}^{n_s} = (X_1,\ldots,X_{n_s}) \in \chi^{n_s}$.

Transmissions are impacted by channel effects and noise that cause the received signal sequence to be perturbed. The complex received vector can be written as 
\begin{align*}
\underline{Y}^{n_s} = (Y_1,\ldots,Y_{n_s}) = \underline{h}^{n_s\times n_s} \underline{X}^{n_s}+\underline{N}^{n_s}, 
\end{align*}
where we assume that the receiver has perfect channel state information (CSI), and so knows both $\underline{h}^{n_s \times n_s}\in\C^{n_s\times n_s}$ and possesses a probabilistic description of $N^{n_s}$, e.g. that it is complex-valued white Gaussian noise with known variance. In Sec. \ref{sec:discussion} we will show to what degree false assumptions about the noise model or CSI impact ORBGRAND-AI's performance.

For ORBGRAND-AI's operation, each received signal corresponding to a coded transmission is split into non-overlapping blocks of $b$ symbols, where for notational ease we assume $n_s/b$ is an integer:
\begin{align*}
\underline{Y}^{n_s} & = \overbrace{
    (Y_1,\ldots,Y_b \,\vert\, \underbrace{ Y_{b+1},\ldots, Y_{2b}}_\text{$b$ symbols} \,\vert\, \cdots \,\vert\, Y_{n_s-b+1},\ldots, Y_{n_s})
    }^\text{$n_s$ symbols} \\
    & = (\underline{Y^b}_1,\ldots,\underline{Y^b}_{n_s/b}).
\end{align*}
Each block $i\in\{1,\ldots,n_s/b\}$ of $b$ symbols, $\underline{Y^b}_i$, is treated separately, with the likelihoods
\begin{align*}
        p_{\underline{X^b}_i|\underline{Y^b}_i}(\underline{t^b}_i|\underline{Y^b}_i) \text{ for each } \underline{t^b}_i\in\chi^b
\end{align*}
being evaluated using the channel model and CSI. We define
\begin{align*}
\underline{t^{b,*}}_i = \arg\max p_{\underline{X^b}_i|\underline{Y^b}_i}(\underline{t^b}_i|\underline{Y^b}_i)
\end{align*}
to be the symbol-level hard demodulation of the block $\underline{Y^b}_i=(Y_{(i-1)b+1},\ldots, Y_{ib})$, which takes channel memory over the block into account. 

The core approximation when evaluating the posterior probability of a noise effect sequence $\underline{t}^{n_s}\in\chi^{n_s}$ describing symbols to be swapped is that the blocks are assumed to be independent, resulting in the following expression
\begin{equation}
    \label{eq:prob_orbgrand_ai}
    \begin{split}
        &p_{\underline{X}^{n_s}|\underline{Y}^{n_s}}(\underline{t}^{n_s}|\underline{Y}^{n_s})\\
        &= \prod_{i=1}^{n_s/b}p_{\underline{X^b}_i|\underline{Y^b}_i}(\underline{\thard}|\underline{Y^b}_i)\prod_{i=1}^{n_s/b}\dfrac{p_{\underline{X^b}_i|\underline{Y^b}_i}(\underline{t^b}_i|\underline{Y^b}_i)}{p_{\underline{X^b}_i|\underline{Y^b}_i}(\underline{\thard}|\underline{Y^b}_i)},
    \end{split}
\end{equation}
which has a common term associated to the sequence of all hard-demodulated blocks and each noise effect sequence that swaps a block experiences a likelihood penalty. 

\begin{algorithm}
\caption{ORBGRAND-AI inputs: The received signal $\underline{Y}^{n_s}$, abandonment threshold $\tau'$, channel statistics $\Psi$ and a codebook membership check function $\codebook$. }
\label{alg:pseudo-code}
\begin{algorithmic}
\STATE {\bf Inputs}: $\underline{Y}^{n_s}$, $\codebook$, $\tau'$, $\Psi$
\STATE {\bf Output}: $c^{n,*}$
\STATE $d'\leftarrow 0$
\STATE $w^\mu\leftarrow$ compute likelihoods for substitution symbol blocks
\WHILE{ $d' < \tau'$}
    \STATE $d'\leftarrow d'+1$ 
    \STATE $e^\mu \leftarrow$ next most likely ORBGRAND pattern for $w^\mu$
    \IF {no substitution conflict}
        \STATE $s^{n_s} \leftarrow$ substitute blocks
        \STATE $c^{n} \leftarrow$ demodulate $s^{n_s}$
        \IF{$\codebook(c^{n}) = 1$}
           \STATE{\bf return} $c^{n,*}\leftarrow c^{n}$
        \ENDIF
    \ENDIF
\ENDWHILE
\STATE{\bf return} FAILURE
\end{algorithmic}
\hrule
\end{algorithm}

With the blocks of symbols, $\underline{t}^i_b$, now playing the role of individual symbols, this expression is identical to the one used for symbol-level ORBGRAND, and so the ORBGRAND approach can be used to generate putative noise effect patterns, $\underline{t}^{n_s}$, in approximately decreasing order of likelihood. In particular, the set of all alternative groups, $\{\underline{t^b}_i\neq \underline{t^{b,*}}_i: i\in\{1,\ldots,{n_s}/b\}\}$, to the hard demodulated blocks of symbols contains $\omega = (2^{m_s b}-1)n/(bm_s)$ elements and they are provided as input to symbol-level ORBGRAND.

To further show that the ORBGRAND pattern generator is well suited to choose substitution blocks, Fig. \ref{fig:subs_likeli} displays the substitution likelihoods of the candidate blocks (block length 4) for various channel conditions at moderate channel correlation $\rho=0.5$ for first order Gauss-Markov noise. Especially at low SNR, the likelihood curve is well approximated by a linear function as assumed by ORBGRAND. For algorithmic clarity, pseudo-code for ORBGRAND-AI can be found in Algorithm \ref{alg:pseudo-code}.

\begin{figure}[htbp]
    \centering
      \begin{subfigure}[b]{0.48\textwidth}
         \centering
         \includegraphics[width=\textwidth]{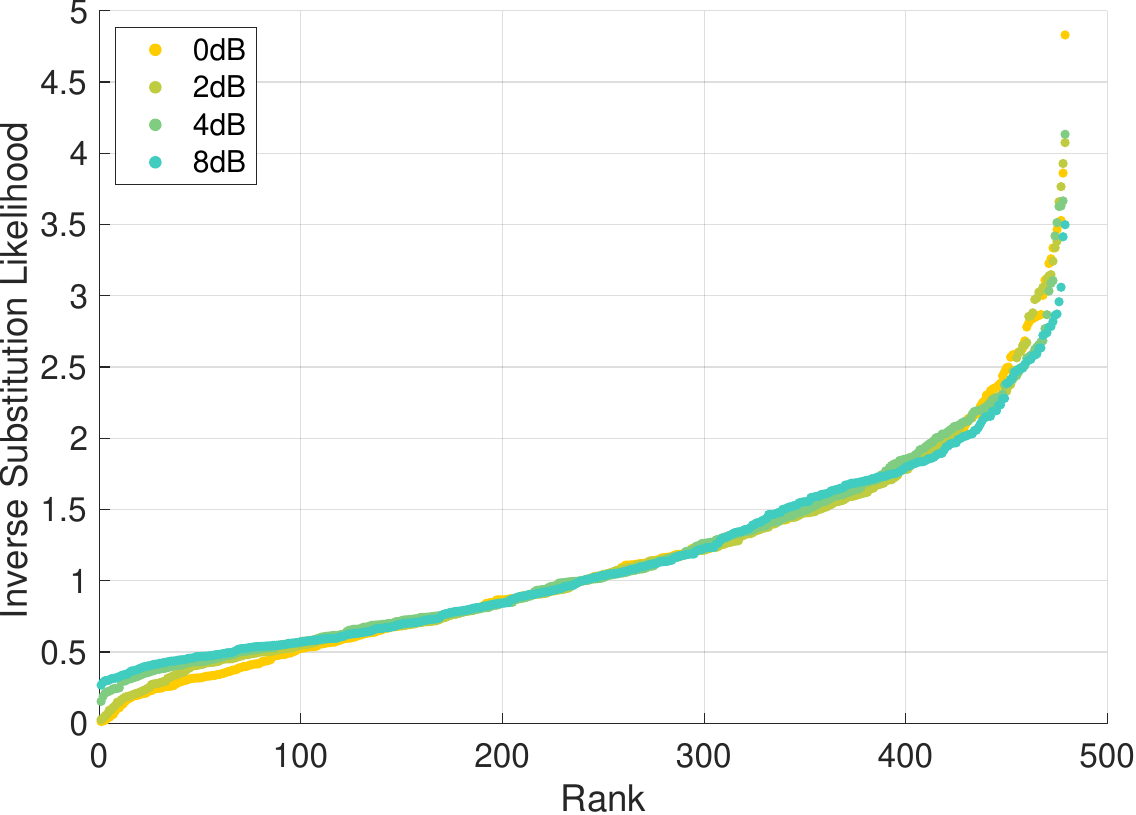}
      \end{subfigure}
    \caption{Ranked normalized inverse block substitution likelihoods for BPSK modulation in a channel with moderately correlated ($\rho=0.5$) first-order Gauss-Markov noise and 128 coded bits. Correlations were only accounted for across blocks of size $b=4$. Blocks with lower rank are more likely to be swapped in.}
    \label{fig:subs_likeli}
\end{figure}

For a Gauss-Markov channel, information theoretic results in Section \ref{sec:capacity} indicate that 
in order to move half-way between the differential entropy rate of the interleaved channel and the differential entropy of the noise with complete correlation, it's sufficient to set $b=2$, suggesting that only small block sizes are necessary to obtain significant performance gains.

\subsection{Illustrative Example}
As an example to show how ORBGRAND-AI constructs code word candidates and queries them iteratively, we use a length $n=4$ code with BPSK ($X_{k'}\in\{+1,-1\})$ and $b=2$. We further assume complex first-order Gauss-Markov noise ($\rho=0.5$, $\sigma^2=1$) resulting in the received symbols $\underline{Y}^{n_s} =[1.5,0.1,-0.2,0.1]^T$. Table \ref{tab:ordered_app} shows the probabilities of the candidate blocks as they appear in eq. \eqref{eq:prob_orbgrand_ai}. Evidently, the most likely (hard detected) symbol vector is $[1,-1,-1,-1]^T$. The remaining blocks are ordered according to their substitution likelihood (Table \ref{tab:ordered_app}). The landslide algorithm can then be used to efficiently generate a substitution order. Table \ref{tab:ordered_subs} displays the order in which the candidate symbol sequences (or rather their corresponding bit representation) will be tested against codebook membership. The first pattern is discarded owing to a substitution conflict that occurs once ORBGRAND tells us to swap substitute 1 and 3, which both belong to the same block, simultaneously (query 6). Note that ORBGRAND-AI could also be deployed with a different pattern generator for combining the blocks in approximately decreasing likelihood order, if desired.
\begin{table}[ht]
    \parbox{.40\linewidth}{
        \centering
        \begin{tabular}{|l||*{5}{c|}}\hline
            \backslashbox{$\underline{X}^2$}{Block}
            &\makebox[1em]{1}&\makebox[1em]{2}\\\hline\hline
                $[+1,+1]$ & 0.30  & 0.38        \\\hline
                $[+1,-1]$ & 0.68  & 0.00        \\\hline
                $[-1,+1]$ & 0       & 0.10        \\\hline
                $[-1,-1]$ & 0.00  & 0.50        \\\hline
        \end{tabular}
    }
    \hfill
    \parbox{.45\linewidth}{
        \centering
        \begin{tabular}{|l||*{5}{c|}}\hline
            \makebox[3em]{Rank}
            &\makebox[3em]{$\underline{X}^2$}&\makebox[3em]{Block}\\\hline\hline
            1 & $[+1,+1]$ & 2        \\\hline
            2 & $[+1,+1]$ & 1        \\\hline
            3 & $[-1,+1]$ & 2        \\\hline
            4 & $[+1,-1]$ & 2        \\\hline
            5 & $[-1,-1]$ & 1        \\\hline
            6 & $[-1,+1]$ & 1        \\\hline
        \end{tabular}
    }
    \caption{Exemplary ORBGRAND-AI demapping procedure result. Left: Probabilities for different block candidates. Right: Ranked (according to their substitution likelihood) substitute blocks.}
    \label{tab:ordered_app}
\end{table}
\begin{table}
    \centering
    \begin{tabular}{|l||*{5}{c|}}\hline
        \makebox[3em]{Query}
        &\makebox[5em]{Swap Indexes}&\makebox[3em]{symbols}\\\hline\hline
        1 & -   & $[+1, -1, -1, -1]$                       \\\hline
        2 & 1   & $[+1, -1, \bm{+1}, \bm{+1}]$             \\\hline
        3 & 2   & $[\bm{+1}, \bm{+1}, -1, -1]$             \\\hline
        4 & 1,2 & $[\bm{+1}, \bm{+1}, \bm{+1}, \bm{+1}]$   \\\hline
        5 & 3   & $[+1, -1 , \bm{-1}, \bm{+1}]$            \\\hline
        6 & 3,1 & discarded                                 \\\hline
        7 & 4   & $[+1, -1, \bm{+1},\bm{-1}]$   \\\hline
    \end{tabular}
    \caption{Illustrative example of a candidate symbol sequence querying order when ORBGRAND patterns are used to pick the blocks to substitute. The first guess corresponds to the hard demodulated sequence. Bold symbols indicate that a swap has taken place.}
    \label{tab:ordered_subs}
\end{table}
\subsection{Higher Order Modulations}
\label{sec:HOM}
While cursory considerations may suggest that ORBGRAND-AI may not be suitable for use with higher order modulations, here we establish that is not the case. For example, using a code with 128 bits, 256-QAM and $b=2$ results in 524,280 potential substitutes for the blocks of the hard demodulated sequence. Similar to symbol level ORBGRAND's approach, we can reduce that number significantly if we only consider symbols in the neighborhood of the corresponding received signal, i.e the $\gamma$ closest symbols. By doing that, we implicitly assume that every substitution candidate which contains a symbol that is not in the neighborhood of its respective received signal is assigned the probability 0. In general, this leads to $(\omega^b-1)n/(m_sb)$ substitution candidate blocks. For $n=128$, 256QAM and a block size $b=2$, we can thus reduce the amount of block substitutes we have to rank order to 120 when choosing $\omega=4$.

\section{Performance Evaluation}
\label{sec:perfeval}
We present the block error rate (BLER) performance of ORBGRAND-AI in the dicode and RFView ISI channels in the following section. In Sec. \ref{sec:perfeval_dicode}, we demonstrate that ORBGRAND-AI can decode any moderate redundancy code and explore the impact of the block size $b$ on the performance gain compared to decoders operating on the interleaved equalized dicode channel, achieving multiple dB gains. We demonstrate the law of diminishing returns for increasing $b$ which was derived in Sec. \ref{sec:capacity}. Finally, in Sec. \ref{sec:rfview_dataset} we show the performance of ORBGRAND-AI in the RFView channel. 

\subsection{Equalized Dicode Channel}
\label{sec:perfeval_dicode}
We have established in Sec. \ref{sec:chan} that first-order Gauss-Markov noise is the result of a two tap channel followed by zero forcing equalization. In the following simulations, we shall refer to the noise power $\sigma_N^2$ as the power of the noise after equalization. This comparison is fair because the benchmark decoders are evaluated on the interleaved channel after equalization.

We first restrict ourselves to BPSK as a modulation scheme. Under these conditions, if we were to drop the interleaver, the performance of CA-Polar codes decoded with CA-SCL decoding would be significantly worse. Fig. \ref{fig:polar} shows that there is an increasing loss as the correlation of the noise samples increases.

\begin{figure}[htbp]
    \centering
      \begin{subfigure}[b]{0.48\textwidth}
         \centering
         \includegraphics[width=\textwidth]{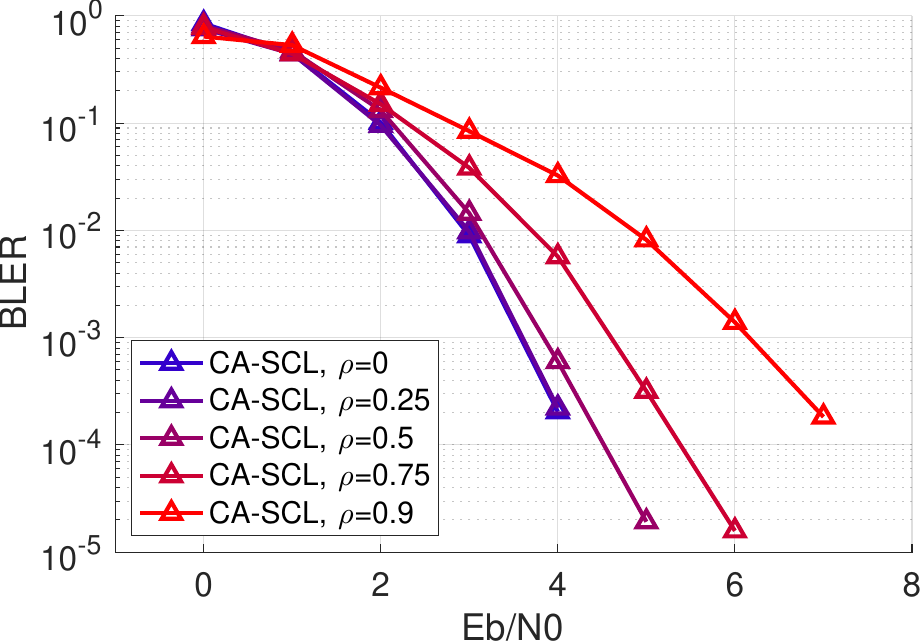}
      \end{subfigure}
    \caption{Impact of channel correlation on CA-SCL decoding of a [128,64] 5G NR CA-Polar code 
    with an 11 bit CRC, list size 8 and within-block interleaver. Block error rate (BLER)
    is plotted versus the energy per information bit, Eb/N0, for the complex equalized dicode channel using BPSK modulation with channel correlation strength $\rho$ that increases from blue to red. 
    }
    \label{fig:polar}
\end{figure}

ORBGRAND-AI on the other hand, operating at a higher rate ([128,110] CA-Polar Code) sees its performance significantly improved with increasing correlation $\rho$ as displayed in Fig. \ref{fig:orbgrand_ai_var_rho}. At a target BLER of $10^{-3}$ ORBGRAND-AI outperforms the interleaved CA-SCL decoder with list size 8 by 2 or even 4dB for $\rho=0.5$ and $\rho=0.75$ respectively. Even at medium correlation ($\rho=0.75$), ORBGRAND-AI is capable of delivering the same $10^{-3}$ BLER as the 1/2 rate CA-Polar code can (Fig. \ref{fig:polar}) under ideal conditions at the same amount of energy spent per bit of information.

The intuition developed in Sec.~\ref{sec:capacity} is reinforced by the results shown in Fig.~\ref{fig:orbgrand_ai_var_b}. For fixed channel conditions with $\rho = 0.5$, increasing the block size $b$ yields diminishing returns in performance gain relative to interleaved ORBGRAND.

To further demonstrate that ORBGRAND-AI effectively captures the impact of channel correlation, Fig.~\ref{fig:orbgrand_ai_var_b} also presents the performance of ORBGRAND paired with an ISI-aware soft demapper and decoder pair based on the iterative extrinsic soft information passing between a BCJR demapper \cite{BCJR74} and ORBGRAND with bit level soft output information which has been shown to deliver the state of the art accuracy for general high redundancy short length codes \cite{yuan2025soft,Kizilates26SOGRAND}. To account for the increase in noise power introduced by equalization, we scale the noise power used by the turbo decoder by the factor $1 - \rho^2$, ensuring that the resulting performance remains comparable to the pre-equalization noise levels.

We observe that, for example, at a target BLER of $10^{-3}$, ORBGRAND-AI with $b=4$ already operates within $0.5\,\mathrm{dB}$ of this benchmark, while using a neighborhood size of $b=8$ reduces the gap to approximately $0.1\,\mathrm{dB}$.

We note, however, that while this approach is feasible for the dicode channel, the computational complexity of the BCJR algorithm grows exponentially with the channel memory, i.e., with the number of channel taps. As a result, its application to channels with moderate to long delay profiles becomes computationally infeasible. In contrast, the demapping complexity of ORBGRAND-AI does not scale with the specific structure of the underlying noise correlation allowing for accounting for noise correlation in channels with moderate or long delay lines as we will see in \ref{sec:rfview}. Instead, it is governed primarily by the chosen neighborhood size $b$, which is treated as a fixed design parameter.

\begin{figure}[htbp]
    \centering
      \begin{subfigure}[b]{0.48\textwidth}
         \centering
         \includegraphics[width=\textwidth]{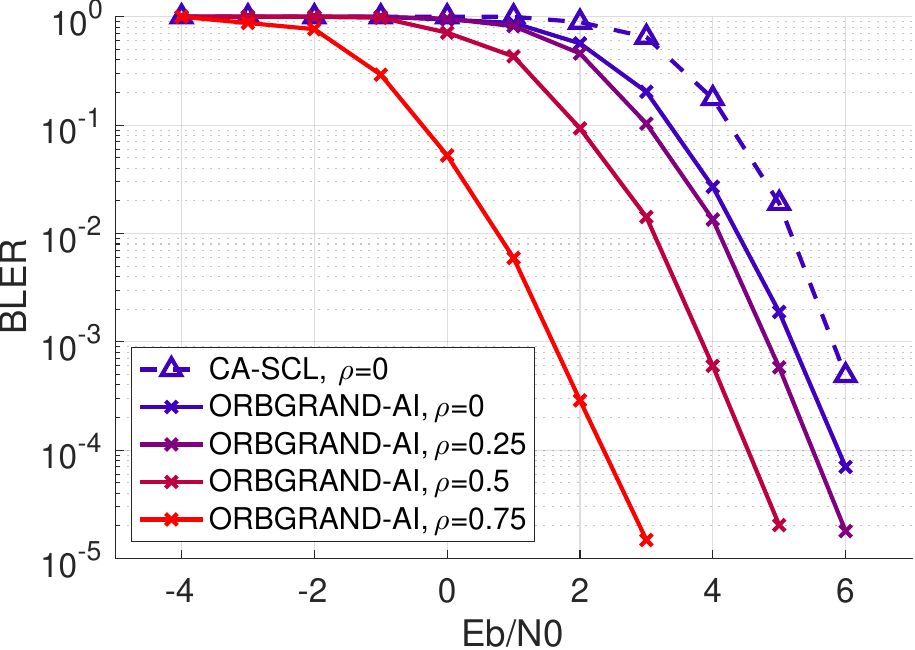}
      \end{subfigure}
    \caption{The impact of channel correlation on ORBGRAND-AI decoding with $b=4$ for a [128,110] 5G NR CA-Polar code with an 11 bit CRC and within-block interleaver for the equalized dicode channel using BPSK modulation. Channel correlation strength, $\rho$, increases from blue to red. The performance of CA-SCL on an interleaved AWGN channel is shown as a benchmark. 
    }
\label{fig:orbgrand_ai_var_rho}
\end{figure}

\begin{figure}[htbp]
    \centering
      \begin{subfigure}[b]{0.48\textwidth}
         \centering
         \includegraphics[width=\textwidth]{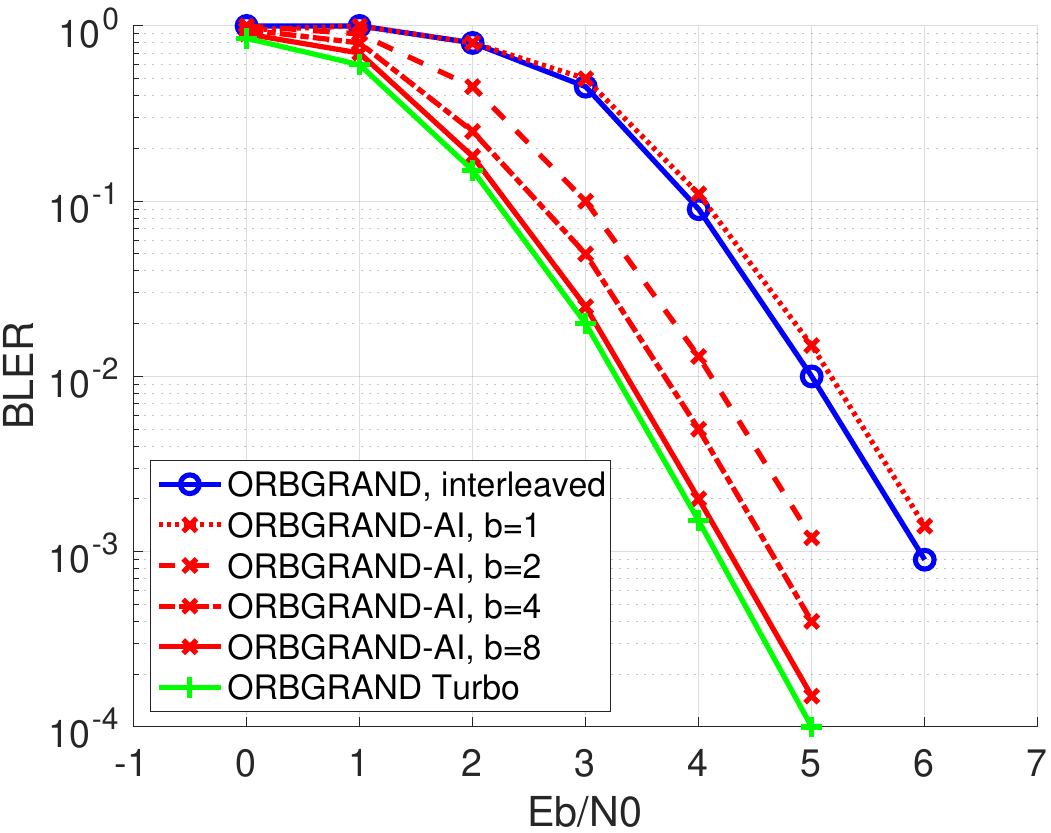}
      \end{subfigure}
    \caption{Impact of the block size, $b$, on the BLER performance of ORBGRAND-AI for $\rho = 0.5$ and a $[128,116]$ RLC over the equalized dicode channel. As a benchmark, we compare against iterative turbo decoding between ORBGRAND and BCJR.}
\label{fig:orbgrand_ai_var_b}
\end{figure}

When we move to higher modulations, as described in Sec. \ref{sec:orbgrand_ai}, we have to make approximations for the sake of complexity reduction. This means that we only consider substitution blocks containing symbols located in the neighborhood of their respective received signals. Fig. \ref{fig:16QAM_ORBGRAND_AI} shows the impact of the neighborhood size $\gamma$ for fixed channel correlation $\rho=0.75$ and block size $b=4$. Obviously, the performance increases when we add more symbols and thus more candidate blocks to the consideration. Further, we see a saturation of performance increase once we consider neighborhoods of 4 or more symbols. This might be due to the fact that in QAM, 4 substitutes suffice to include a potential substitute in every direction in the I-Q plane.

The runtime complexity of ORBGRAND-AI is driven by two coupled components: correlation-aware soft-information processing, including reliability generation, adaptation to the local correlation structure, and reliability ordering; and the GRAND search itself, whose core operations are noise pattern generation and codebook membership checking. For GRAND-family decoders, efficient hardware architectures for reliability-driven ordering and on-the-fly pattern generation have already been demonstrated, e.g. \cite{abbas2022high,condo2022fixed,Riaz23,xiao2023low,riaz2024sub}.

A decoder-internal complexity measure that is commonly considered for GRAND algorithms is the average number of codebook queries required until decoding succeeds, since the associated search process typically dominate the latency and energy consumption of resulting chip implementations \cite{riaz2024sub}. This metric provides a proxy for the internal complexity of the GRAND decoding stage, although it does not by itself capture the full receiver cost, which also includes soft-information generation, correlation adaptation, and sorting. Figure~\ref{fig:queries_16QAM_ORBGRAND_AI_duplicate} reports this average query count for ORBGRAND-AI. For a fixed neighborhood size \(\gamma=5\), ORBGRAND-AI with block size \(b=4\) requires approximately 100 codebook queries on average at a target BLER of \(10^{-4}\), indicating operation in a low average-search regime under typical conditions.

To limit the worst-case complexity of GRAND, practical implementations employ query abandonment, whereby the decoder terminates the search once a predefined query budget has been reached \cite{duffy2019grand,riaz2021multicode,riaz2025sub}. This provides an explicit upper bound on the worst-case decoding latency and energy consumption. If no valid codeword has been identified within the allowed budget, the decoder declares an erasure or decoding failure. The query budget can be selected such that this worst-case bound is enforced at negligible cost to decoding performance, as successful decoding events are dominated by the most likely noise patterns, which occur early in the GRAND query sequence \cite{duffy2019grand,riaz2025sub}.

With regard to a more holistic assessment, we point to recently published taped out ASIC implementation of ORBGRAND-AI that incorporates on-chip soft demapping, substitution pattern generation and candidate construction, and codebook checking \cite{kizilates2025low}. Results therein provide strong evidence that ORBGRAND-AI is not only algorithmically attractive, but also practically realizable in low-latency and energy-efficient hardware implementations.

\begin{figure}[htbp]
    \centering
    \centering
      \begin{subfigure}[b]{0.48\textwidth}
         \centering
         \includegraphics[width=\textwidth]{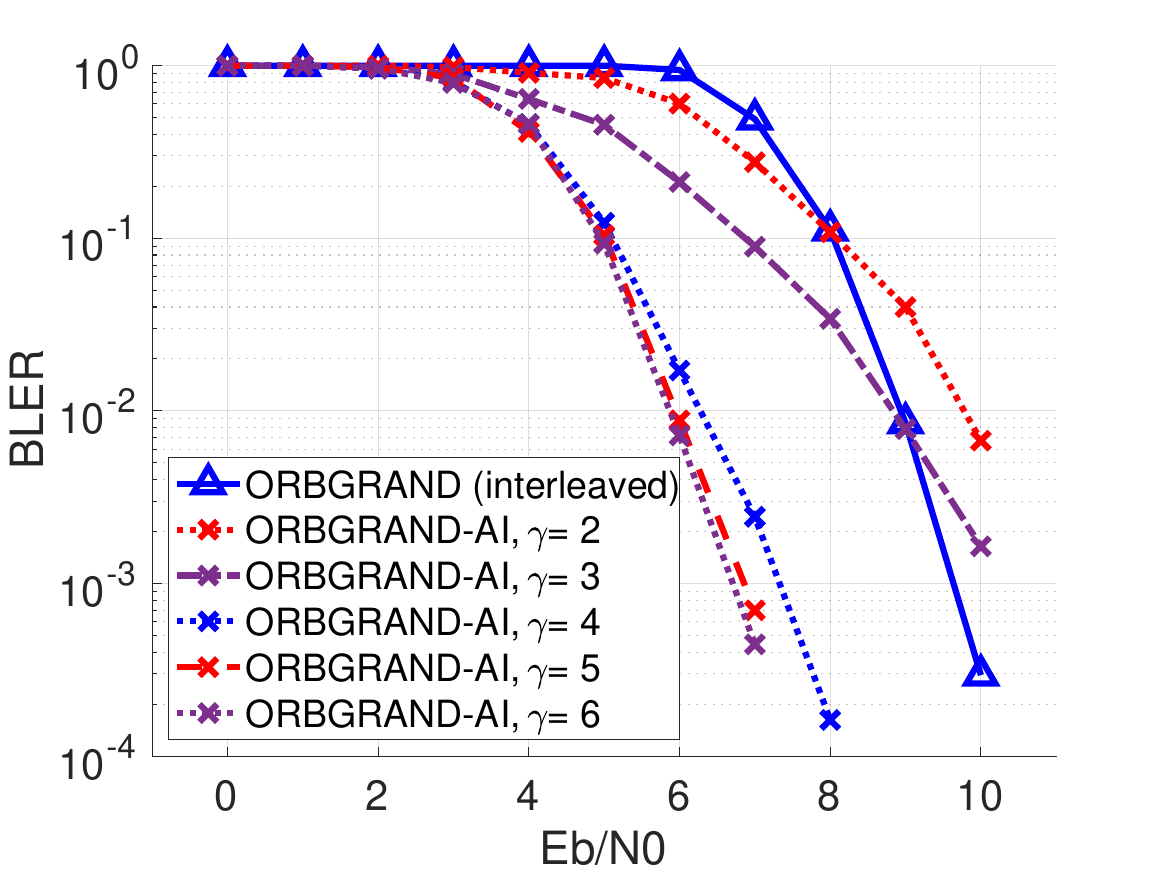}
      \end{subfigure}
    \caption{[256,240+11] 5G NR Uplink CA-Polar code with 11 bit CRC, 16QAM, with $\rho=0.75$ in the complex equalized dicode channel. The block size is fixed at $b=4$ taking into account $\gamma$ candidates per symbol. }
    \label{fig:16QAM_ORBGRAND_AI}
\end{figure}

\begin{figure}[htbp]
    \centering
      \begin{subfigure}[b]{0.48\textwidth}
         \centering
         \includegraphics[width=\textwidth]{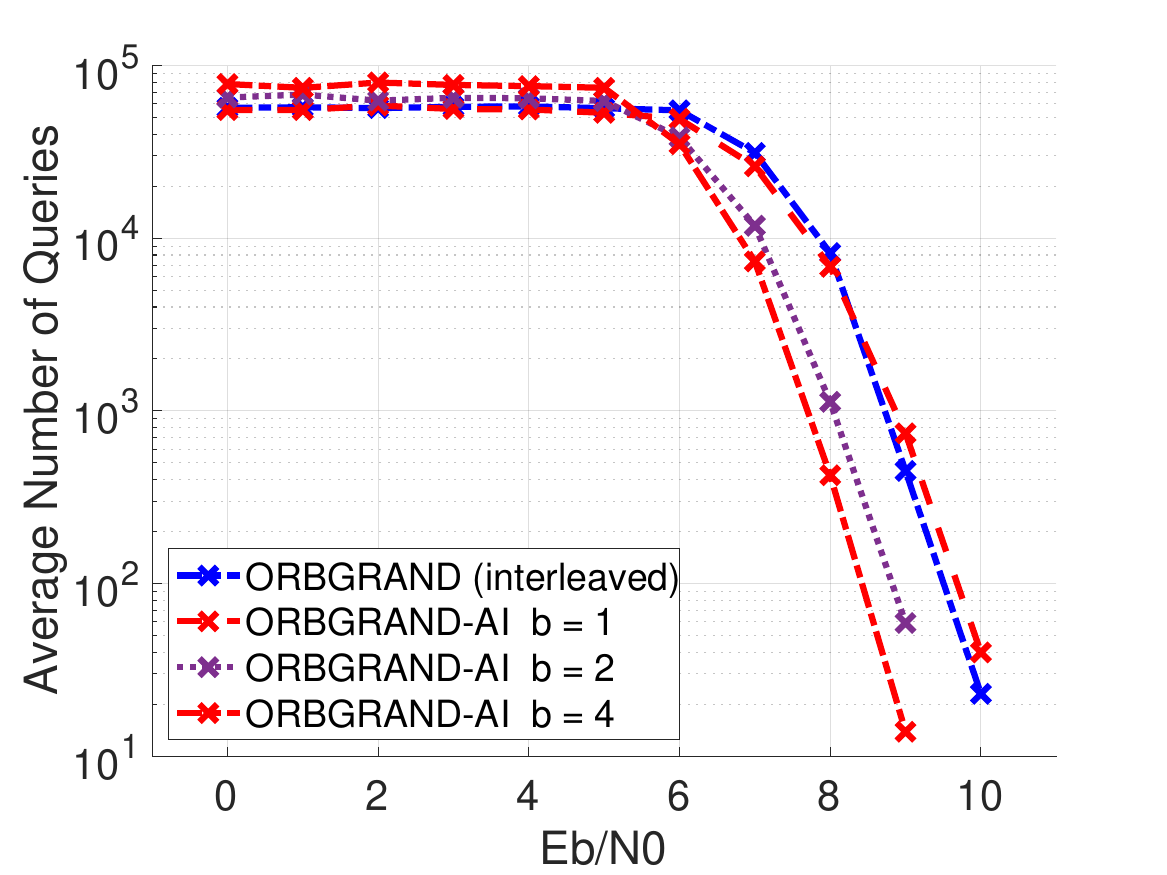}
      \end{subfigure}
    \caption{Number of codebook queries it takes ORBGRAND-AI until a code word is found for $\rho=0.5$ and an [256, 240+11] CA-Polar code in the equalized dicode channel. Only $\gamma=5$ substitutes were considered per 16QAM modulated symbol.}
    \label{fig:queries_16QAM_ORBGRAND_AI_duplicate}
\end{figure}

\subsection{Equalized ISI Channel from RFView Dataset}\label{sec:rfview_dataset}
\label{sec:rfview}

Finally,  we investigate the performance of ORBGRAND-AI in the RFView channel, $\underline{h}^{n_s\times n_s}_{ISI}$, as described in Sec. \ref{sec:rfview_dataset_channel}. In these simulations $\sigma_N^2$ denotes the variance of $\underline{N}^{n_s}$ before equalization. To decode, $\underline{X}^{n_s}$, the received symbols, $\underline{Y}^{n_s}$, are equalized using a minimum mean-square error (MMSE) equalizer: 
\begin{align*}
    \underline{h}_{\text{eq, MMSE}}^{n_s\times n_s} =& \underline{C}^{n_s \times n_s
}_{X} (\underline{h}_{ISI}^{n_s\times n_s})^H ( \underline{h}_{ISI} \underline{C}^{n_s \times n_s
}_{X} (\underline{h}_{ISI}^{n_s \times n_s})^H \\& + \underline{C}^{n_s \times n_s
}_{N})^{-1}
\end{align*}
\noindent where $\underline{C}^{n_s \times n_s
}_{X}$ and $\underline{C}^{n_s \times n_s
}_{N}$ denote the auto-covariance matrices of $\underline{X}^{n_S}$ and $\underline{N}^{n_s}$ respectively and the operator $(\cdot)^H$ denotes the Hermitian transpose. The equalized symbols are denoted by
\begin{align*}
    \underline{Y_{\text{eq}}^{n_s}} &= \underline{h}_{\text{eq, MMSE}}^{n_s\times n_s} \underline{Y}^{n_s}\\
    &= \underline{h}_{\text{eq, MMSE}}^{n_s \times n_s} (\underline{h}^{n_s\times n_s}_{ISI} \underline{X}^{n_s} + \underline{Y}^{n_s})
\end{align*}
\noindent The equalized channel output $\underline{Y^{n_s}_{eq}}$ is provided to the GRAND decoder. 

The auto-covariance matrix of the equalized symbols, $\underline{C}^{n_S\times n_s}_{Y_{eq}}$, provides the colored noise statistics to the GRAND decoder where: 
\begin{align} \label{covariance_matrix}
    \underline{C}^{n_s\times n_S}_{Y_{\text{eq}}} &= \mathbb{E}[(\underline{Y_{eq}^{n_s}} - \underline{X}^{n_s})
    (\underline{Y_{eq}^{n_s}} - \underline{X}^{n_s})^H] \nonumber \\
    &=\underline{h}^{n_s \times n_s}_{\text{eq, MMSE}} \underline{h}^{n_s \times n_s
}_{ISI} \underline{C}^{n_s \times n_s
}_{X} (\underline{h}^{n_s \times n_s
}_{ISI})^H (\underline{h}^{n_s \times n_s
}_{\text{eq, MMSE}})^H  \nonumber \\
    & \ \ \ +\underline{h}^{n_s \times n_s
}_{\text{eq, MMSE}} \underline{C}^{n_s \times n_s
}_{N} (\underline{h}^{n_s \times n_s
}_{\text{eq, MMSE}})^H  \nonumber \\
    & \ \ \ -\underline{h}^{n_s \times n_s
}_{\text{eq, MMSE}} \underline{h}^{n_s \times n_s
}_{ISI} \underline{C}^{n_s \times n_s
}_{X} \nonumber \\
& \ \ \ -\underline{C}^{n_s \times n_s
}_{X} (\underline{h}^{n_s \times n_s
}_{ISI})^H (\underline{h}^{n_s \times n_s
}_{\text{eq}})^H + \underline{C}^{n_s \times n_s
}_{X}, 
\end{align}
\noindent that is in Algorithm \ref{alg:pseudo-code}, $\Psi = \underline{C}_{Y_{eq}}^{n_s\times n_s}$. In the ORBGRAND-AI algorithm, correlation over small blocks of symbols is considered. To compute $\underline{C}_{Y_{eq}}^{n_s\times n_s}$ for a particular block, we use the following covariance matrix 
\begin{align*}
    \underline{C}^{n_s \times n_s
}_{X|\underline{X}^{b}_i} = \mathbb{E}[&[X_1... X_{ib} ... X_{ib + (b-1)}...X_n]^H\cdot\\
    &[X_1 ... X_{ib} ... X_{ib+(b-1)}... X_n]]
\end{align*}

\noindent in eqn.~(\ref{covariance_matrix}).

The equalization process colors the noise in the channel. This means that the equalized channel output which is an estimate of the transmitted BPSK symbol is observed in colored noise. This is reflected in the auto-covariance matrix, $\underline{C}^{n_s \times n_s}_{Y_{\text{eq}}}$, which has non-zero off-diagonal elements.

Figure \ref{fig:isi_bler} shows that by accounting for the coloring in the noise due to ISI over 2.5 dB gain  can be obtained in BLER performance using $b = 2$. BLER performance improves as the block size increases because colored, or correlated, channels have lower entropy and therefore have higher capacity \cite{Cover91}. Figure \ref{fig:isi_bler} also shows that by adding a forward error correcting code a further 4 dB gain can be obtained in terms of BLER. These results show that by applying forward error correction coding and accounting for any coloring, or statistical correlation, in the channel over a 6 dB improvement BLER can be obtained over uncoded systems.

\begin{figure}[htbp]
    \centering
      \begin{subfigure}[b]{0.48\textwidth}
         \centering
         \includegraphics[width=\textwidth]{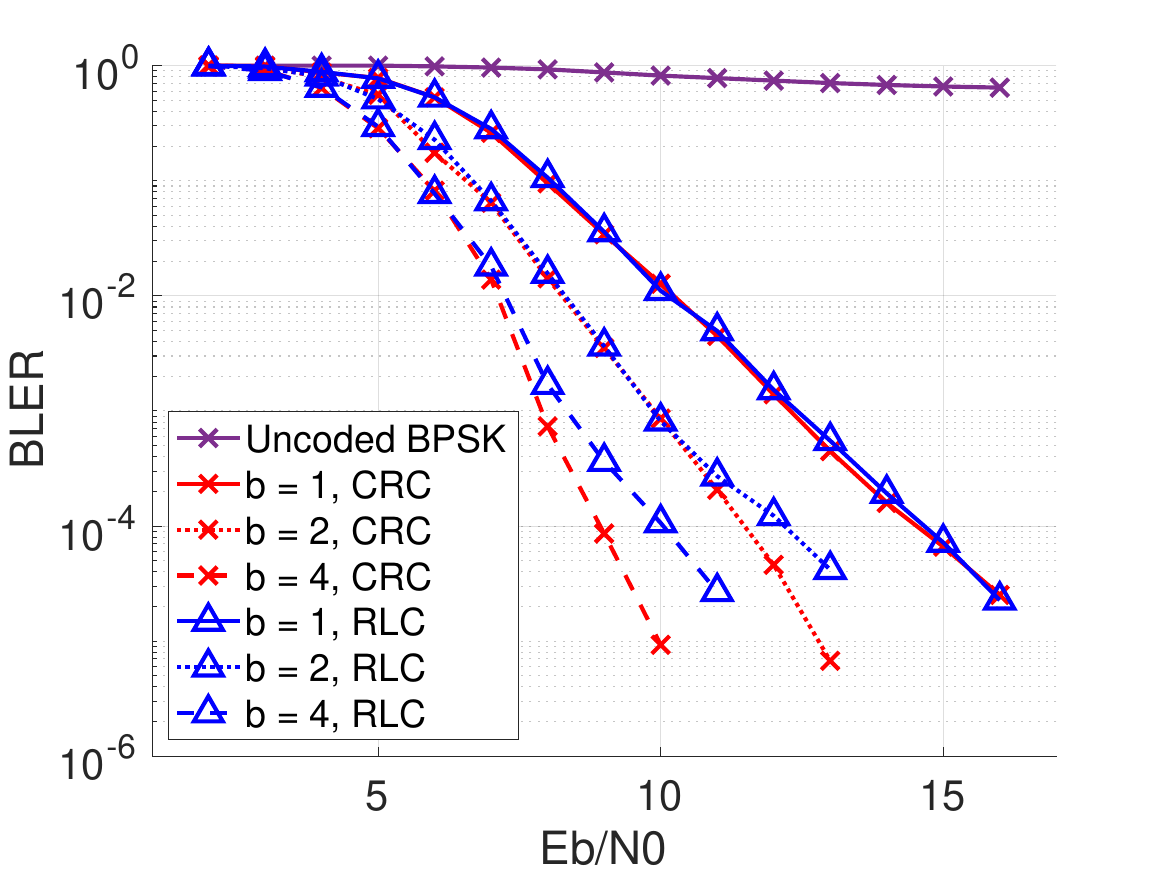}
      \end{subfigure}
    \caption{ Comparison of BLER for different block sizes, $b$, for a [132,120] random linear code and [132,120] cyclic-redundancy  code with Koopman polynomial 0xb41 \cite{koopmancode} MMSE equalization and ORBGRAND-AI decoding in the equalized RFView ISI channel, $\underline{h}_{RFV}$.}
    \label{fig:isi_bler}
\end{figure}

\section{Robustness Considerations} \label{sec:discussion}

In practice, CSI is always subject to error \cite{tse2005fundamentals}. This is mostly due to the fact that precise channel estimation methods are costly in terms of the number of pilot symbols needed and thus there is a trade off between the percentage of pilot symbols used and the quality of CSI \cite{coleri2002channel}. Another cause of imperfection is quantization of parameters used in the processing of the received signal. Hence, decoders are desired to be as robust to mismatched CSI as possible. In the case of ORBGRAND-AI, this question translates to how imperfect CSI impacts the query order of noise sequences. In the following sections we consider the effect of measurement and quantization error on decoding performance in both the dicode and RFView channels. 

\subsection{Equalized Dicode Channel}

For the equalized dicode channel, we first explore the effect of measurement error. We assume the  estimation error to be additive and normally distributed, $h_{k, est} = h_k(1 + \epsilon_k)$, where the variance of $\epsilon_k$ is known as normalized mean squared error (NMSE). In fact, the estimation error not only impacts the query order of ORBGRAND-AI, but also leads to incorrect equalization resulting in an error floor. Fig. \ref{fig:incorr_csi} shows the result of the equalized dicode channel for various NMSE values. For a significant NMSE of $0.1$, the error floor is clearly visible. 

To further isolate the effect of a mismatch in the decoder, Fig. \ref{fig:orbgrand_ai_missmatch_rho} displays the degradation of ORBGRAND-AI's performance for a quantization error $\Delta\rho$ regarding $\rho$ in the decoding process for $\rho_{real}=0.5$ where $\rho = \rho_{real} + \Delta\rho$. We see that for a considerable mismatch of even $\Delta\rho=0.2$, the performance degradation is still less than $0.5\text{dB}$. A reason for the query order's robustness against imperfect CSI may lie in the fact that for the ordering of potential substitutes, the exact strength of statistical correlation between neighboring symbols is not as important as the fact that they are correlated at all.

\begin{figure}[htbp]
    \centering
      \begin{subfigure}[b]{0.48\textwidth}
         \centering
         \includegraphics[width=\textwidth]{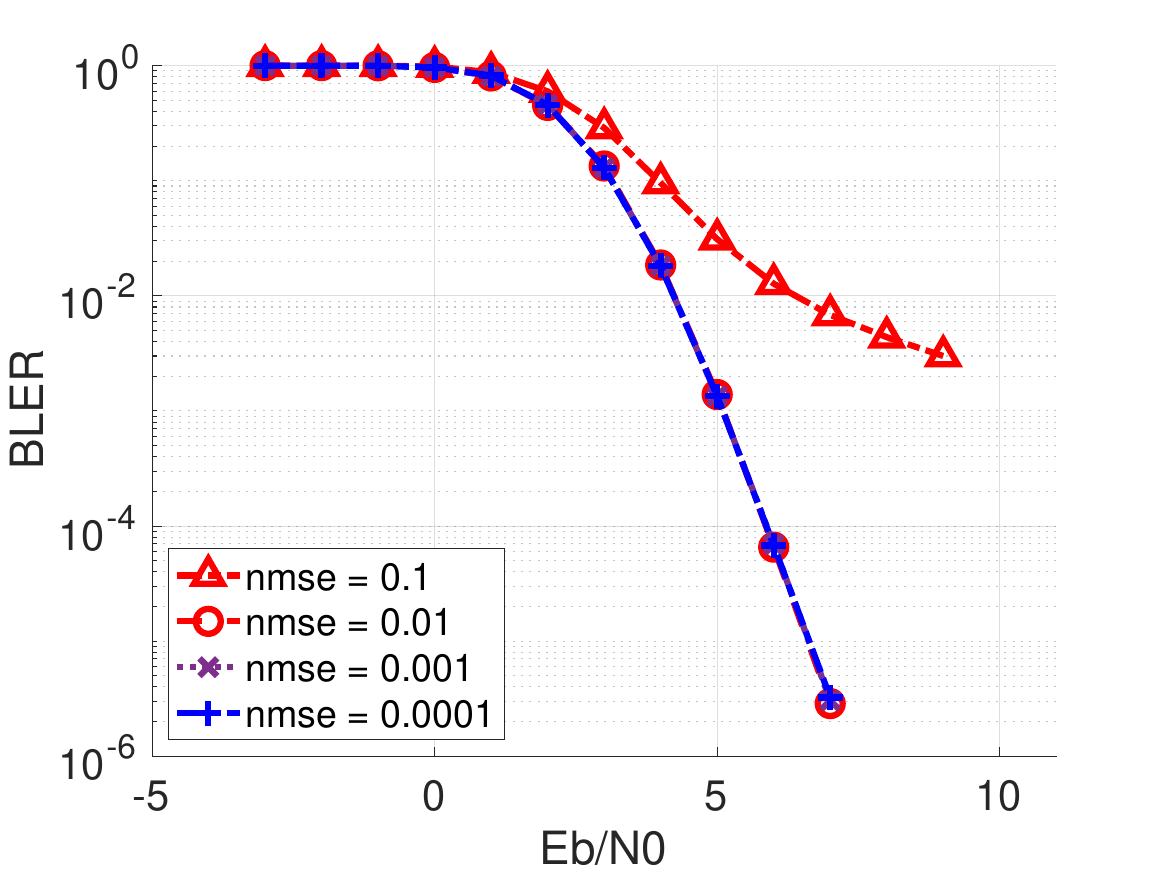}
      \end{subfigure}
    \caption{ ORBGRAND-AI performance degradation due to imperfect equalization in the equalized dicode channel with $\rho=0.75$ and normally distributed additive estimation error with variance $\text{nmse}$. We used a [128, 116] RLC and BPSK with $b = 4$.}
    \label{fig:incorr_csi}
\end{figure}

\begin{figure}[htbp]
    \centering
      \begin{subfigure}[b]{0.48\textwidth}
         \centering
         \includegraphics[width=\textwidth]{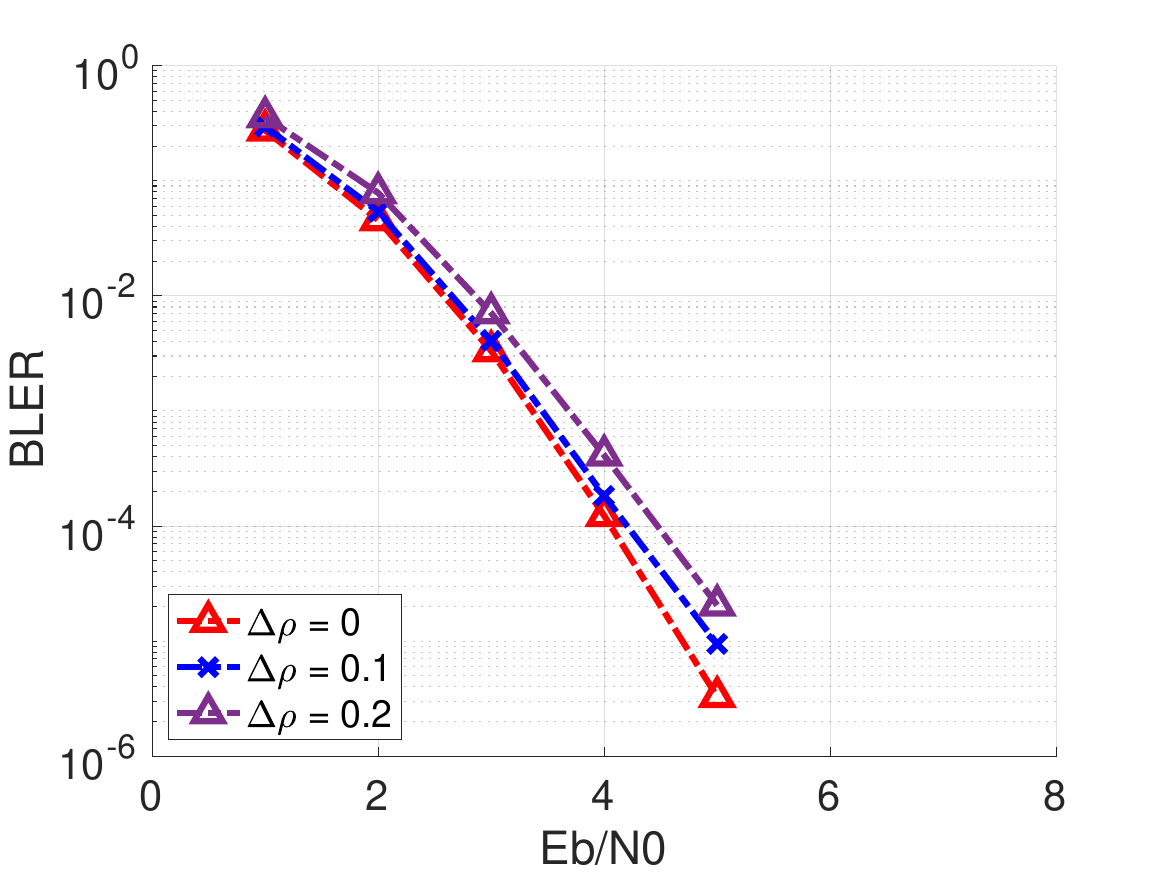}
      \end{subfigure}
    \caption{ORBGRAND-AI's sensitivity to a CSI quantization error $\Delta\rho$ in an equalized dicode channel with $\rho_{real}=0.5$. A (128,112) CRC with polynomial $0x9eb2$ \cite{koopmancode} was used alongside BPSK and a fixed block size $b=4$.}
    \label{fig:orbgrand_ai_missmatch_rho}
\end{figure}

\subsection{Equalized ISI Channel from RFView Dataset}

We first investigate the effect of measurement error in the RFView channel by approximating the RFView channel as a second order autoregressive (AR(2)) process.
To do this, we fit an AR(2) process to each set of channel coefficients from the matched filter output $\underline{z}''^{6, k'}$ for each sounding signal $k'$. We denote the resulting matrix of channel coefficients obtained from the AR(2) approximation by $\underline{\hat{h}}^{n_s\times n_s}_{AR(2)}$. We selected an AR(2) process because the performance of a first order autoregressive (AR(1)) process estimate was poor. The AR(1) channel estimate quickly diverges from the true estimate yielding poor BLER performance. 

We wish to model the matched filter output $\underline{z}''^{6, k'}$  as an AR(2) process, i.e. we wish to find coefficients $\phi_1$ and $\phi_2$ such that
\begin{equation*}
    z_{j', k'} = \phi_1 z_{j'-1, k'} + \phi_2 z_{j'-2,k'} + \epsilon_{j'}
\end{equation*}
for $j' \in [1,...6]$ and where $\epsilon_{j'}$ denotes the Gaussian innovation process. We calculate estimates $\hat{\phi}_1$ and $\hat{\phi}_2$ of $\phi_1$ and $\phi_2$ respectively using the least squares estimate for each sounding signal $k'$
\begin{equation*}
    [\hat{\phi}_{1, k'} \ \hat{\phi}_{2, k'}]^T = ((\underline{\bar{z}}_{k'}^{4\times 2})^H\underline{\bar{z}}^{4\times 2}_{k'})^{-1}(\underline{\bar{z}}^{4\times2}_{k'})^H\underline{\tilde{z}^4_{k'}}. 
\end{equation*}
where 
\begin{equation*}
    \underline{\bar{z}}^{4\times 2}_{k'} = \begin{bmatrix} \underline{z}''_{1, k'} & \underline{z}''_{2, k'} \\ 
    \vdots & \vdots \\
    \underline{z}''_{4, k'} & \underline{z}''_{5, k'} \end{bmatrix} 
\end{equation*}
and 
\begin{equation*}
    \underline{\tilde{z}}^{4}_{k'} = [\underline{z}''_{3, k'} \dots \underline{z}''_{6, k'}]^T.
\end{equation*}
\noindent We use $\phi_1$ and $\phi_2$ to denote the AR(2) parameters instead of $\rho_1$ and $\rho_2$ as was done in section \ref{sec:capacity} because when we compute the least squares estimates $\hat{\phi}_1, \hat{\phi}_2$ we are not guaranteed to obtain numbers in the range $(0,1)$.

Given initialization conditions $\hat{z}''_{1, k'} = z''_{1, k'}$ and $\hat{z}''_{2, k'} = z''_{2, k'}$, and using $\hat{\phi}_{1, k'}$ and $\hat{\phi}_{2, k'}$, we approximate the remaining four coefficients in the matched filter output as 
\begin{equation*}
    \hat{z}_{j', k'} = \hat{\phi}_{1,k'} \hat{z}_{j'-1, k'} + \hat{\phi}_{2,k'} \hat{z}_{j'-2,k'}
\end{equation*}
for $j > 2$. Now we can construct $\underline{\hat{h}}^{n_s\times n_s}_{AR(2)}$ using the sampling process outlined previously by sampling from $\underline{\hat{z}}''^{6, k'}$ instead.

We now use the AR(2) estimate, $\underline{\hat{h}}^{n_s\times n_s}_{AR(2)}$, of the channel in the equalization and covariance matrix calculations in place of $\underline{h}^{n_s\times n_s}_{RFV}$.  In Fig.~\ref{fig:isi_ar2_bler}, we compare the BLER of the performance of ORBGRAND-AI with perfect CSI and imperfect CSI with the channel approximated with an AR(2) process. We observe that there is high concordance between the case with perfect CSI and the case where we only have access to an AR(2) estimate of the channel. This shows that even with imperfect channel estimates we are able to obtain performance gains with ORBGRAND-AI.

\begin{figure}[htbp]
    \centering
      \begin{subfigure}[b]{0.48\textwidth}
         \centering
         \includegraphics[width=\textwidth]{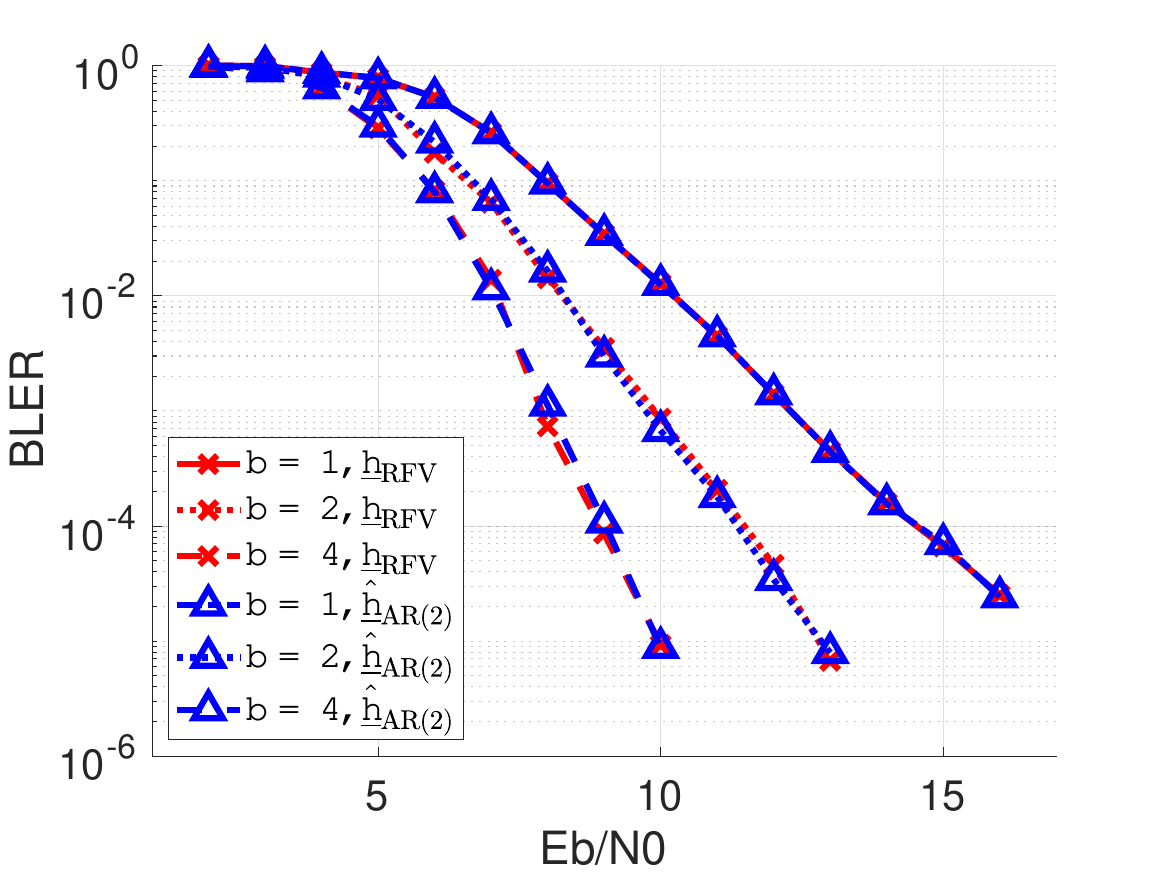}
      \end{subfigure}
    \caption{Comparison of BLER for different block sizes, $b$, using a [132,120] cyclic-redundancy code with Koopman polynomial 0xb41 \cite{koopmancode} using MMSE equalization with perfect CSI, $\underline{h}^{n_s \times n_s}_{RFV}$, and the AR(2) process approximation of the RFView channel, $\underline{\hat{h}}^{n_s \times n_s}_{AR(2)}$ using ORBGRAND-AI decoding.}
    \label{fig:isi_ar2_bler}
\end{figure}

Next, we investigate the effect of quantization on the RFView channel. To quantize the RFView channel, we find the minimum and maximum of both the real and imaginary components of all channel coefficients in $\underline{h}^{n_s\times n_s}_{RFV}$. Then, we create $q'$ evenly spaced quantization levels between both the maximum and minimum of the real and imaginary components of the channel coefficients. We then map the original channel coefficients to their quantized counterparts represented by the matrix $\underline{\hat{h}}^{n_s\times n_s}_{q'}$. Fig.~\ref{fig:isi_quantized_25_v2} shows that when we do not take into account channel correlation for $q' = 25$, i.e. when $b = 1$, the BLER performance under the quantized scheme plateaus at high $E_b/N_0$. At high $E_b/N_0$ values, the noise introduced by the quantization scheme becomes the dominant source of error. As we consider correlation by increasing the block size, $b$, for $q' = 25$ we find that we recover BLER performance, therefore showing that we can mitigate the effects of quantization noise by accounting for the correlation. For a higher quantization level of $q' = 100$ we observe performance similar to the perfect CSI case.

\begin{figure}[htbp]
    \centering
      \begin{subfigure}[b]{0.48\textwidth}
         \centering
         \includegraphics[width=\textwidth]{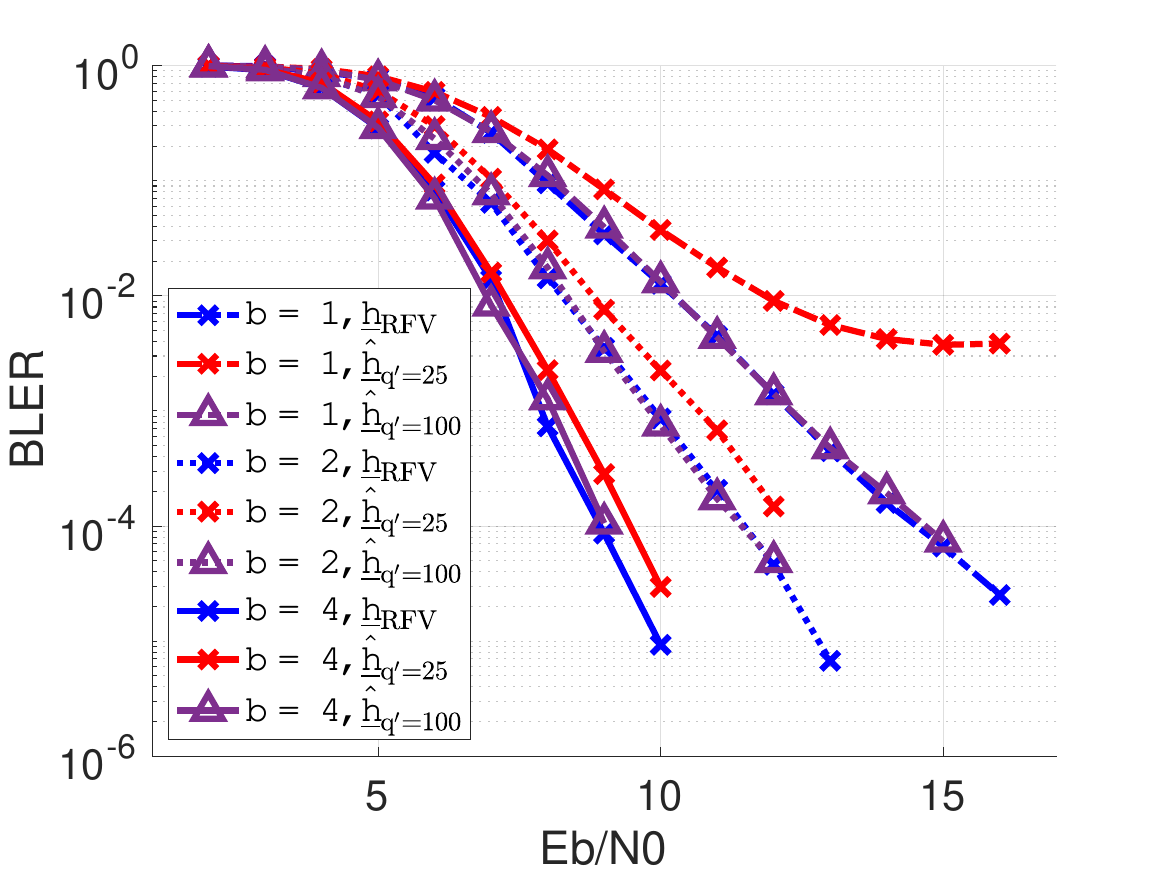}
      \end{subfigure}
    \caption{Comparison of BLER for different block sizes, $b$, using a [132,120] cyclic-redundancy code with Koopman polynomial 0xb41 \cite{koopmancode} MMSE equalization with perfect CSI, $\underline{h}^{n_s\times n_s}_{RFV}$, and the 25 and 100-level quantization of the RFView channel, $\underline{\hat{h}}^{n_s\times n_s}_{q'=25}$ and $\underline{\hat{h}}^{n_s\times n_s}_{q'=100}$ respectively, using ORBGRAND-AI decoding.}
    \label{fig:isi_quantized_25_v2}
\end{figure}

\section{Conclusion}

We have presented ORBGRAND-AI, a decoding framework that explicitly accounts for temporal correlation in channel-induced noise without turbo equalization. We demonstrate that significant improvements in both block error rate (BLER) and bit error rate (BER) can be achieved when the decoding process directly exploits the correlation structure of the noise rather than suppressing it through interleaving. These gains are consistently observed across multiple datasets, channel realizations, and operating conditions.

Through systematic evaluation, we show that ORBGRAND-AI maintains strong performance under practical implementation constraints. In particular, the decoder proves robust to channel estimation inaccuracies and quantization of its input parameters, indicating that the benefits of correlation-aware decoding persist even in practical receiver implementations.

These results suggest an alternative design for modern communication transceivers. Rather than using fine-grained interleavers to artificially remove correlation in order to match decoder assumptions or an iterative process of equalization and decoding, ORBGRAND-AI directly exploits this structure during decoding. This opens a pathway toward enabling ultra-reliable low-latency communication.

Finally, some natural extensions of this work merit further investigation. First, a study of the interaction between ORBGRAND-AI and channel sensing, as considered in \cite{millward2024enhancing}. As channel sensing introduces correlated measurement errors, this structure could potentially be exploited to devise sensing strategies that are better matched to ORBGRAND-AI decoding. Second, a theoretical analysis of error exponents under the approximate independence assumption would provide further insight into the fundamental performance limits of the proposed approach. 

\bibliographystyle{IEEEtran}
\bibliography{grand,ORBGRAND,moritz, rfview}

\end{document}